\documentclass[a4paper,11pt,nohyper]{JHEP3}
\usepackage{amsmath,latexsym,amssymb}
\usepackage{graphicx}
\usepackage[latin1]{inputenc}
\usepackage{subfigure}
\usepackage{nicefrac}

\usepackage{comment}

\newcommand\nn{\nonumber}
\newcommand\beal{\begin{align}}

\title{AdS$_\mathbf{4}$ flux vacua in type II superstrings and their domain-wall solutions}

\author{Costas Kounnas${}^{\diamondsuit}$, Dieter Lüst${}^{\clubsuit\heartsuit}$,
 P.~Marios~Petropoulos${}^{\spadesuit}$ and Dimitrios Tsimpis${}^{\clubsuit}$ \\

  \begin{itemize}
  
\item Laboratoire de Physique Théorique de l'Ecole Normale Supérieure, CNRS -- UMR 8549\\
      24 rue Lhomond, 75231 Paris Cedex 05, France

\item  Arnold-Sommerfeld-Center for Theoretical Physics\\
Department für Physik, Ludwig-Maximilians-Universität München\\
Teresienstra\ss e 37, 80333 München, Germany

\item  Max-Planck-Institut für Physik\\
Föhringer Ring 6, 80805 München, Germany

  \item  Centre de Physique Théorique, Ecole Polytechnique, CNRS -- UMR 7644 \\
    91128 Palaiseau, France
  \end{itemize}

\bigskip
 E-mail:
\email{kounnas@lpt.ens.fr, luest@theorie.physik.uni-muenchen.de,
marios@cpht.polytechnique.fr,
tsimpis@theorie.physik.uni-muenchen.de}}

\abstract{We investigate the emergence of supersymmetric
negative-vacuum-energy ground states in four dimensions. First, we
rely on the analysis of the effective superpotential, which
depends on the background fluxes of the internal manifold, or
equivalently has its origin in the underlying gauged supergravity.
Four-dimensional, supersymmetric anti-de Sitter vacua with all
moduli stabilized appear when appropriate Ramond and
Neveu--Schwarz fluxes are introduced in IIA. Geometric fluxes are
not necessary. Then the whole setup is  analyzed from the
perspective of the sources, namely D/NS-branes or Kaluza--Klein
monopoles. Orientifold planes are also required for tadpole
cancellation. The solutions found in four dimensions correspond to
domain walls interpolating between $\mathrm{AdS}_4$ and flat
spacetime. The various consistency conditions (equations of
motion, Bianchi identities and tadpole cancellation conditions)
are always satisfied, albeit with source terms. We also speculate
on the possibility of assigning (formal) entropies to
$\mathrm{AdS}_4$ flux vacua via the corresponding dual brane
systems. }

\preprint{ LPTENS/07/32\\
  LMU-ASC 50/07\\
   MPP-2007-98\\
  CPTH-RR060.0707
  }

\keywords{anti-de Sitter vacua, fluxes, branes}

\begin{document}

\setcounter{footnote}{0}
\renewcommand{\thefootnote}{\arabic{footnote}}
\setcounter{section}{0}
\section{Introduction}\label{sec:intro}

The closed/open string correspondence between flux vacua and
D-brane solutions has led to remarkable new insights on the
structure of gauge and gravity theories. Anti-de Sitter
(AdS) geometries play a crucial role in this context since they
appear on one side of the AdS/CFT duality
\cite{Maldacena:1997re,Witten:1998qj,Gubser:1998bc}. The latter was
originally formulated between $\mathrm{AdS}_5$ extended
supergravity and ${\cal N}=4$ $SU(N)$ super Yang--Mills on the
boundary of $\mathrm{AdS}_5$. It is based on the observation that
the type IIB, $\mathrm{AdS}_5\times S^5$ ground state with
non-vanishing Ramond 5-form flux arises as the near-horizon
geometry of a stack of D3-branes with open strings attached.
Another example with maximal supersymmetry is ${\cal N}=8$
supergravity on $\mathrm{AdS}_{4(7)}\times S^{7(4)}$ in
eleven-dimensional supergravity with non-vanishing 4-form flux.
Again, these $\mathrm{AdS}$ background spaces arise as
near-horizon geometries of M2(5)-branes.

In fact, the holographic correspondence relies on the possibility
of viewing $N$ units of 5-form flux in ten-dimensional
supergravity as originating from a stack of $N$ D3-branes. This
interplay between flux backgrounds and branes is richer than the
simple relation that exists between a field and the sources that
create it and they will play a central role in the present note.
Indeed, the sources at hand are truly dynamical objects that bring
their own degrees of freedom and, depending on the regime, the
latter may become important ingredients.

Anti-de Sitter geometries arise also in string backgrounds with
fewer supersymmetries, such as $\mathrm{AdS}_5\times T^{1,1}$,
which arises from placing a stack of D3-branes near the
six-dimensional conifold singularity \cite{Klebanov:1998hh}. This ${\cal N}=2$
supersymmetric background is dual to a conformal ${\cal N}=1$,
$SU(N)\times SU(N)$ gauge theory with bifundamental matter fields.
The geometric transitions \cite{Vafa:2000wi} are another nice example of the
flux/brane correspondence where one trades fluxes through internal
cycles for their dual sources made of wrapped D-branes. In this
way an intriguing relation emerges between the confining phase in
${\cal N}=1$ effective gauge theories and their microscopic
$SU(N)$ descriptions. Moreover, string backgrounds of the form
$\mathrm{AdS}_3\times S^3\times T^4(K3)$ can be constructed by
conformal field theory techniques and play an important role in
the context of dualities between three-dimensional gravity and
two-dimensional boundary CFT's \cite{Maldacena:1998bw,Kounnas:2000nd}.

The investigation of the flux/brane correspondence in conjunction
with the generated anti-de Sitter backgrounds is also
motivated by the possibility of comparing the macroscopic and
microscopic entropies of supersymmetric black holes \cite{Strominger:1996sh,Callan:1996dv}.
Supersymmetric black holes in ${\cal N}=2$ type IIA/B supergravity \cite{Ferrara:1995ih}
arise as solutions with non-vanishing electric/magnetic charges of
$U(1)$ Ramond gauge fields in four dimensions. Their near-horizon
geometry is given by the space $\mathrm{AdS}_2\times S^2$. The
corresponding thermodynamic entropies are determined by the area
of the horizon plus corrections from higher-curvature terms in the
effective action \cite{Behrndt:1996jn,Maldacena:1997de, Lopes Cardoso:1998wt}. The same $\mathrm{AdS}_2\times S^2$ geometry is
obtained by adopting the brane-source picture.

The sources are in
this case intersecting D-branes, point-like in four dimensions,
and hence with the spatial part of their world-volumes entirely inside the internal
six-dimensional part of space: D3-branes wrapped around internal
Calabi--Yau (CY) 3-cycles in type IIB, or D4-branes wrapped around
4-cycles plus D0-branes in type IIA. Counting the open-string
excitations living on the wrapped branes, the resulting
microscopic entropy matches up the macroscopic energy of the black
holes in the effective supergravity theory, as demonstrated in
several explicit CY examples. These observations are far reaching
and have led to many subsequent developments: wave functions of
certain supersymmetric $\mathrm{AdS}_2\times S^2$ flux vacua \cite{Ooguri:2005vr},
attractor formalism and generalized non-supersymmetric black holes (for a recent review
of black holes in string theory see \cite{Mohaupt:2007mb}).
%

Clearly, much more important and phenomenologically interesting
than two-dimensional $\mathrm{AdS}_2$ flux vacua  is to consider
four-dimensional $\mathrm{AdS}_4$ supergravity solutions, which will be the
subject of this paper. This
raises a threefold problem: (\romannumeral1) provide stable
supersymmetric $\mathrm{AdS}_4$ flux vacua within supergravity
theory, (\romannumeral2) identify the corresponding brane pictures
and (\romannumeral3) compute the entropy by counting the
microscopic string states of the dual branes.
In this paper we try to take the first two steps in this program
by deriving supersymmetric, intersecting brane configurations
generating four-dimensional $\mathrm{AdS}_4$ flux vacua.


Our framework will be perturbative type II supergravity. We will
in particular concentrate on type IIA theory with non-vanishing
Ramond, Neveu--Schwarz, and possibly geometrical fluxes that can
in principle generate $\mathrm{AdS}_4$ ground states without using
further non-perturbative effects. Despite the power of
gauged-supergravity tools that can be used here, this exercise is
not simple because the requirements of unbroken supersymmetry,
with negative vacuum energy, and without runaway or flat
directions are not easy to fulfill simultaneously. Type IIB
solutions with similar properties can also be obtained. All this
is the content of Sec. \ref{fluco}.

The next step consists in describing the above $\mathrm{AdS}_4$
flux vacua as near-horizon geometries of appropriately distributed
intersecting D2/D4/D6/D8/NS5-branes and Kaluza--Klein monopoles.
The examination of these, as well as of certain T-dual IIB
systems, will be undertaken in Sec. \ref{ex}. Except for exotic
non-geometric setups briefly described in Sec. \ref{2.3}, full
stabilization is hard to achieve perturbatively in type IIB CY
compactifications because of the Kähler-moduli independence of the
superpotential. Nevertheless, we will introduce a new and
interesting type IIB model by appropriately distributing
D3/D5/D7/NS5/KK-branes/monopoles. Kaluza--Klein monopoles are
sources for geometrical fluxes induced by the Scherk--Schwarz
mechanism \cite{ftss}\footnote{From a string-theory viewpoint, the
Scherk--Schwarz mechanism is implemented as a freely acting
orbifold \cite{strss}.}. They introduce torsion and thereby
Kähler-moduli dependence. In the present case all moduli are
stabilized around an $\mathrm{AdS}_4$ vacuum, which appears as the
near-horizon geometry of the above intersecting extended sources.
The corresponding internal six-dimensional geometry is now a
\emph{nilmanifold}.

The general treatment of the source-branes, including the subtle
issue of Bianchi identities and tadpole cancellation conditions is
 discussed in generality in Sec. \ref{replacing}.  It is
made clear there that some of the $\mathrm{AdS}_4$ vacua derived
from a four-dimensional effective superpotential, {\it do not}
admit a ten-dimensional supergravity interpretation with
\emph{localized} brane/orientifold sources. Indeed, as we will see
in Sec. \ref{2.3}, from the point-of-view of the four-dimensional
effective superpotential there exist supersymmetric
$\mathrm{AdS}_4$ flux vacua \emph{without} metric fluxes.
However, such compactifications do not exist in the context of
ten-dimensional supergravity \cite{Lust:2004ig}, unless one
considers the addition of {\it smeared} brane and/or orientifold
sources \cite{acha}. The latter are precisely the ingredients
needed for evading the necessity of metric fluxes. We note in passing
that the smearing can be thought of as an artifact of the
supergravity approximation, which necessarily ignores the massive
Kaluza--Klein modes.

Let us finally mention that from the four-dimensional perspective,
the branes act as two-branes, i.e. \emph{domain walls}. This is now
perfectly consistent, as opposed to Minkowski vacua, since the
geometry is not flat but interpolates between $\mathrm{AdS}_4$ and
asymptotically flat space. Related considerations, including
domain-wall solutions
in the context of (gauged) supergravity, have previously
appeared in
\cite{Cvetic:1991vp}-\cite{mk}.

The domain-wall picture is very useful,
since the tension of the domain walls can be viewed as the origin
of the effective ${\cal N}=1$ superpotential. Specifically, in
this context one has to solve the supersymmetry conditions and
equations of motion for the domain wall coupled to the various
scalar fields. We expect that in such a solution the scalar fields
would be non-singular in spacetime, since in the corresponding
flux vacua they are all fixed to finite values. We therefore
believe that for the supersymmetric $\mathrm{AdS}_4$ domain-wall
solutions there should exist an attractor mechanism that
determines the values of the scalar fields at the $\mathrm{AdS}_4$
horizon -- in analogy to the attractor mechanism for the
supersymmetric $\mathrm{AdS}_2\times S^2$ black holes in ${\cal
N}=2$ supergravity. To explicitly prove this statement is beyond
the scope of this work.

\section{AdS$_\mathbf{4}$ vacua from flux compactifications}
\label{fluco}

In the following we will exhibit type IIA/B models with Ramond,
Neveu--Schwarz or geometrical fluxes that possess supersymmetric
$\mathrm{AdS}_4$ ground states. We will focus on perturbative
(tree-level) contributions and use the effective superpotential
description in four dimensions, omitting possible contributions to
the scalar potential by D-terms.

\subsection{Searching for minima}\label{genmin}

In a general ${\cal N}=1$ supergravity, the superpotential
$W(\phi)$ is a function of chiral superfields $\phi_i$. The
corresponding scalar potential takes the standard form
\begin{equation}\label{scapot}
V=\mathrm{e}^K\left(|D_iW|^2-3|W|^2\right)\, ,
\end{equation}
where the F-terms are defined as
\begin{equation}\label{fterm}
F_i=\mathrm{e}^{K/2}D_iW=\mathrm{e}^{K/2}\left(
\partial_{\phi_i}W+W\partial_{\phi_i}K\right)
\end{equation}
with $K$ being the Kähler potential.

Our aim is to find supersymmetric extrema of $V$. We must
therefore impose
\begin{equation}\label{ftermmin}
F_i(\phi_{\mathrm{min}})=0 \ \forall  i\, .
\end{equation}
Anti-de Sitter vacua require negative vacuum energy. Equations
(\ref{scapot}) and (\ref{ftermmin}) thus lead to the following
requirement:
\begin{equation}\label{Wnzer}
W(\phi_{\mathrm{min}})\neq 0\, .
\end{equation}

Generically, conditions (\ref{ftermmin}) and (\ref{Wnzer}) are not
easy to satisfy. One might assume e.g.
$\partial_{\phi_i}K\vert_{\mathrm{min}}\neq 0$ for the main moduli
fields. Using (\ref{ftermmin}) and (\ref{Wnzer}), this translates
into $\partial_{\phi_i}W\vert_{\mathrm{min}}\neq 0$, which in turn
shows that $W$ must depend on all main moduli fields $\phi_i$. By this
reasoning, one would therefore conclude that supersymmetry
together with negative vacuum energy require full moduli
dependence of $W$ and imply the stabilization of \emph{all}
moduli.\footnote{Strictly speaking, the real parts of all moduli
are stabilized.} However, the assumption on non vanishing
$\partial_{\phi_i}K\vert_{\mathrm{min}}$ does not hold generically
(except for the seven main moduli considered in this paper) and
supersymmetric $\mathrm{AdS}_4$ do not systematically ensure
all-moduli stabilization. This issue is more subtle and deserves
therefore a careful analysis, which is one of our motivations
here. Fortunately, this is somewhat easier that the corresponding
search for Minkowskian vacua with (un)broken supersymmetry and
with all moduli fixed.

\subsection{Fluxes in type IIA/B and mirror map}\label{2.2}

\paragraph{Type IIB} We will start without geometrical fluxes; then the tree-level 3-form flux
superpotential in type IIB on a Calabi--Yau 3-fold $X$ is of the
standard form \cite{Gukov:1999ya,Taylor:1999ii,Mayr:2000hh,Curio:2000sc,Giddings:2001yu}. It gets two kinds
of contributions, namely from Ramond and Neveu--Schwarz 3-form
fluxes:
\begin{eqnarray}\label{tv}
W_{\mathrm{IIB}}&=&W_H+W_F=\int_X\Omega\wedge \left( F^{\mathrm{R}}_3+SH^{\mathrm{NS}}_3\right)\nonumber \\
&=&e_0+ie_iU_i+i m_0F_0(U)+m_iF_i(U)\nonumber
\\ &&+iS(a_0+ia_iU_i+i b_0F_0(U)+b_iF_i(U))\, .
\end{eqnarray}
Here $\Omega$ is the holomorphic 3-form on the CY space, and
$F^{\mathrm{R}}_3$ ($H^{\mathrm{NS}}_3$) is the Ramond
(Neveu--Schwarz) 3-form field strength field. The $U$-dependent
function $F(U)\equiv F_0(U)$ is the holomorphic prepotential and
the $F_i(U)$ are its first derivatives. The $e_I,m_I$ comprise the
Ramond 3-form fluxes, whereas the $a_I,b_I$ correspond to the
Neveu--Schwarz 3-form fluxes ($I=0,\dots , h^{2,1}$). The
superpotential $W$ depends on the complex-structure moduli fields
$U_i$ ($i=1,\dots ,h^{2,1}$) and on the dilaton $S$, whereas it is
independent of the K\"ahler moduli $T_m$ ($m=1,\dots ,h^{1,1}$).

In type IIB the fluxes generate a $C_4$ tadpole given by
\begin{equation}\label{iibtadpole}
N_{\rm flux}=\int H_3\wedge F_3 =
\sum_{I=0}^{h^{2,1}}a_Im_I+b_Ie_I\, .
\end{equation}
This flux number is equivalent to the Ramomd charge of $N_{\rm
flux}$ D3-branes, and has to be cancelled by the  orientifold
O3-planes and an appropriate number of D3-branes.

To be more precise, type II compactification on CY threefolds
results in $\mathcal{N}=2$ supergravity in four dimensions. One
may then introduce orientifolds which reduce the four-dimensional
supersymmetry to $\mathcal{N}=1$. The orientifold involution
introduces a grading on the spaces of harmonic forms of the CY,
under which each cohomology group $H^n$ splits into even/odd
subspaces with respect to the involution:  $H^n=H^n_+\oplus
H^n_-$. Accordingly, the dimension of the K\"{a}hler moduli space
of type IIA/IIB is given by $h^{1,1}_{\mp}$; similarly the
dimension of the complex-structure moduli space is truncated to
$h^3_{\pm}=1+h^{2,1}$. With this understanding,
 we will omit the $\pm$ subscript on the various CY cohomology groups.
We refer the reader to \cite{gl} for a detailed discussion of CY
orientifold compactifications.

According to our previous discussion (Sec. \ref{genmin}), for
geometrical CY spaces with $h^{1,1}>0$, since $\partial_{T_m}W=0$,
it follows that supersymmetry is not compatible with negative
vacuum energy.  However it is compatible with the flat no-scale
models with (un)broken supersymmetry. This holds under the
assumption that $\partial_{\phi_i}K \neq 0$, which is certainly
fulfilled for the Kähler moduli. Supersymmetric $\mathrm{AdS}_4$
vacua in type IIB compactifications with all moduli stabilized
seem therefore unlikely, unless one adds a non-perturbative
superpotential that depends on the K\"ahler moduli $T_m$.

The problem of the Kähler-moduli perturbative independence can be
in principle resolved either by introducing geometrical fluxes,
i.e. torsion, hence abandoning the CY structure, or in backgrounds
which are non-geometrical and do not possess \emph{at all}
K\"ahler moduli (i.e. $h^{1,1}=0$ in the framework of CY). These
may be thought of as asymmetric orbifolds, or as the mirror of the
$Z$-manifold, and have been recently discussed in a different
context \cite{Becker:2007dn}\footnote{It is fair to say that the
existence of such compactifications in type IIB is still
conjectural. It is inferred by T-duality from type IIA, where a
precise geometric construction is available, based on the
identification of all complex-structure moduli with $S$. This
construction was also discussed some time ago in
\cite{Candelas:1993nd}.}. In this case there are no associated
F-term conditions, and supersymmetric $\mathrm{AdS}_4$ can be
obtained in type IIB directly from the tree-level flux
superpotential (Eq. (\ref{tv})) with $S$ and all complex-structure
moduli $U_i$ stabilized (see Sec. \ref{2.3} for the detailed
computation on the type IIA side). Concerning the models with
torsion, a new example will be discussed in Sec. \ref{D3D5}.

\paragraph{Type IIA}

The moduli dependence is different in type IIA compactifications.
Even in the absence of geometrical fluxes, the problem of the
Kähler-moduli perturbative independence does not generically
occur. Situations of this type, involving only Neveu--Schwarz and
Ramond fields, will be examined later. Furthermore, in most cases,
the inclusion of geometrical fluxes is not an option but becomes
compulsory. Indeed, in type IIA the backreaction due to the Ramond
and Neveu--Schwarz fluxes is usually so strong that the underlying
geometry $Y$ is no longer a CY space, but rather a space with a
certain $SU(3)$ or $SU(2)$ group structure. These spaces have
generically $\mathrm{d}J\neq 0$, $J$ being the K\"ahler form,
which leads precisely to geometrical-flux
contributions\footnote{These geometrical fluxes originate from the
mirror transformation (fibre-wise T-duality) of the Neveu--Schwarz
3-form fluxes $a_i$ on the type IIB side.}. Those reinforce the
Kähler-moduli dependence of the perturbative effective
superpotential.

Notice that there are nonetheless examples where not all moduli
are present. The type IIA mirror of the Kähler-moduli-free example
cited above has no complex-structure moduli (in the CY language
this corresponds to $\tilde h^{2,1}=0$). Although this is not a
priori necessary for the perturbative stabilization of all moduli,
we will discuss this particular scenario in Sec. \ref{2.3} because
stabilization occurs in this case thanks to the particular
structure of the Kähler potential.

The type IIA effective superpotential receives three kinds of
contributions (see e.g.
\cite{Derendinger:2004jn,Villadoro:2005cu,DeWolfe:2005uu,Camara:2005dc}):
\begin{equation}\label{IIAfull}
W_{\mathrm{IIA}}=W_H+W_F+W_{\mathrm{geom}}\, .
\end{equation}
The first term is due to the Neveu--Schwarz 3-form fluxes and
depends on the dilaton $S$ and the type  IIA complex-structure
moduli $U_m$ ($m=1,\dots ,\tilde h^{2,1}$):
\begin{equation}W_H(S,U)=\int_Y\Omega_c\wedge H_3=i\tilde a_0S+i\tilde c_mU_m\, ,
\end{equation}
where in type IIA the 3-form $\Omega_c$ is defined by
$\Omega_c=C_3+i\mathrm{Re}(C\Omega)$. Second, we have the
contribution from Ramond 0-, 2-, 4-, 6-form fluxes (the 0-form
flux corresponds to the mass parameter $\tilde m_0$ in massive IIA
supergravity -- see also \cite{Lust:2004ig}):
\begin{eqnarray}
W_F(T) &=&\int_Y \mathrm{e}^{J_c}\wedge F^{\mathrm{R}}\nonumber
\\ &=&\tilde m_0\frac{1}{6}\int_Y\left(J_c\wedge J_c\wedge
J_c\right) +\frac{1}{2}\int_Y\left(F_2^{\mathrm{R}}\wedge
J_c\wedge J_c\right)+
\int_YF_4^{\mathrm{R}}\wedge J_c+\int_YF_6^{\mathrm{R}}\nonumber \\
&=&i\tilde m_0F_0(T)-\tilde m_iF_i(T)+i\tilde e_iT_i+\tilde
e_0\,\label{IIAF} .
\end{eqnarray}
Here $F(T):= F_0(T)$ is the type IIA prepotential, which depends
on the IIA K\"ahler moduli $T_i$ ($i=1,\dots ,\tilde h^{1,1}$) and
$F_i(T):=\partial F_0/\partial T_i$. We use the notation $J_c$ for
the complexified K\"{a}hler metric $J_c:=B+iJ$. Finally we have
the contribution of the geometrical (metric) fluxes, which
captures the non-Calabi--Yau property of $Y$:
\begin{eqnarray}
W_{\mathrm{geom}}(S,T,U)=i\int_Y\Omega_c\wedge \mathrm{d}J=-\tilde
a_iST_i-\tilde d_{im}T_iU_m\, ,\label{IIAg}
\end{eqnarray}
where the metric fluxes $\tilde a_i$, $\tilde d_{im}$ parameterize
the non-vanishing of $\mathrm{d}J$.

The type IIA Ramond tadpole follows from the equation of motion of
the field $C_7$. Specifically it is of the form
\cite{Camara:2005dc}:
\begin{equation}\label{iiatadpole}
\tilde N_{\rm flux}=\int\left(C_7\wedge
\mathrm{d}F_2+C_7\wedge(\tilde a_0H_3+\mathrm{d}\bar F_2)\right)\,
,
\end{equation}
where $G_2=\mathrm{d}C_1+\tilde a_0B_2+\bar F_2$ and
$^*F_2=F_8=\mathrm{d}C_7$. The metric fluxes $\tilde a_i$
contribute to $\mathrm{d}\bar F_2$, and one gets for non-vanishing
fluxes $\tilde a_I$ and $\tilde m_I$ that
\begin{equation}\label{naflux}
\tilde N_{\rm flux}=\sum_{I=0}^{\tilde h^{1,1}}\tilde a_I\tilde
m_I\, .
\end{equation}
This non-vanishing flux tadpole, which corresponds to a
non-vanishing D6-brane charge,  must be cancelled by the orientifold O6-planes and an appropriate
number of  D6-branes:
\begin{equation}
\label{tadpoleiia1} \tilde N_{\rm
flux}+N_{\mathrm{D6}}=2N_{\mathrm{O6}}\, .
\end{equation}

\paragraph{The mirror map}

Let us now assume that the two spaces $X$ and $Y$ are mirror
(T-)dual to each other. This implies that $h^{1,1}=\tilde
h^{2,1}$, $h^{2,1}=\tilde h^{1,1}$. Moroever we identify
$S^{\mathrm{IIA}}=S^{\mathrm{IIB}}$,
$T^{\mathrm{IIA}}=U^{\mathrm{IIB}}$,
$U^{\mathrm{IIA}}=T^{\mathrm{IIB}}$. Then we can see that on the
various fluxes in the type IIA/B superpotential mirror symmetry
acts as follows. On the one hand, all Ramond fluxes are mapped
onto each other, i.e.
\begin{equation}\label{emmatch}
e_I=\tilde e_I\, ,\quad m_I=\tilde m_I\, .
\end{equation}

On the other hand, Neveu--Schwarz 3-fluxes in type IIB are
generically mapped on metric fluxes in type IIA and vice versa.
The precise mapping is determined by the orientation of the $T^3$
in $X$ on which the T-duality is acting. We assume that the IIB
Neveu--Schwarz 3-form flux $a_0$ remains a Neveu--Schwarz 3-form
flux in type IIA, whereas the Neveu--Schwarz 3-form fluxes $a_i$
become metric fluxes. Specifically we get
\begin{equation}\label{amatch}
a_I=\tilde a_I\, .
\end{equation}

The type IIB Neveu--Schwarz 3-form fluxes $b_I$ as well as the
type IIA Neveu--Schwarz 3-form fluxes $\tilde c_m$ and metric
fluxes $\tilde d_{im}$ do also have their mirror-duals. We will
not elaborate any further on the precise correspondence,
but mention that tadpoles (Eq. (\ref{naflux})) do also match when
using the full mirror dictionary (including (\ref{emmatch}) and
(\ref{amatch})).

\subsection{Supersymmetric AdS$_\mathbf{4}$ vacua in type IIA and their IIB mirror duals}
\label{2.3}

We now come to the crucial point of generating supersymmetric
$\mathrm{AdS}_4$ ground states in type IIA with all main moduli
stabilized. One of the few instances where such vacua were
constructed is provided in \cite{Derendinger:2004jn}. The present
construction is somewhat different and we will comment on this
later, in particular in Sec. \ref{replacing}.

The situation we will consider falls in the class where the
assumption $\partial_{\phi_i}K\vert_{\mathrm{min}}\neq 0$ holds.
Therefore as emphasized in Sec. \ref{genmin} the superpotential
must depend on all chiral fields for the vacuum energy to be
negative with unbroken supersymmetry. Following
\cite{Derendinger:2004jn, DeWolfe:2005uu, Camara:2005dc} we will
concentrate on the case without metric fluxes, i.e. $\tilde
a_i=\tilde d_{im} =0$. Furthermore, the 6-form fluxes as well as
the 2-form fluxes can be shown to be gauge dependent and hence can
be set to zero: $\tilde e_0=\tilde m_i=0$ \cite{DeWolfe:2005uu,
Camara:2005dc}. The fluxes $\tilde m_0$ and $\tilde a_0$ must be
non-zero for $W$ to be kept non-vanishing. Finally, we combine
$\tilde e_i$ and $\tilde m_0$ as
\begin{equation}
\gamma_i=\tilde m_0\tilde e_i\, .
\end{equation}

Assuming a simple (toroidal) cubic prepotential $F=T_1T_2T_3$, the
superpotential has the generic form:
\begin{eqnarray}
W_{\mathrm{IIA}}&=&W_F+W_H= \tilde m_0\int_Y(J\wedge J\wedge J)+
\int_YF_4^{\mathrm{R}}\wedge J+
\int_Y\Omega_c\wedge H_3\nonumber\\
&=&i\tilde e_iT_i+i\tilde m_0T_1T_2T_3+i\tilde a_0S+i\tilde
c_mU_m\, .\label{IIAU}
\end{eqnarray}
Using Eq. (\ref{naflux}), the D6-tadpole of corresponding fluxes
is simply given by
\begin{equation}\label{nafluxa}
\tilde N_{\rm flux}=\tilde a_0\tilde m_0\, .
\end{equation}
According to Eq. (\ref{tadpoleiia1}) this number has to be
balanced by the D6-branes and the O6-planes.

We may also consider the generalization to the case where the
prepotential is given by $F=\frac{1}{6}c_{ijk}T_iT_jT_k$. In CY
compactifications, the $c_{ijk}$ would be the classical triple
intersection numbers and the corresponding superpotential would
read:
\begin{eqnarray}
W_{\mathrm{IIA}}&=&W_F+W_H =i\tilde e_iT_i+i\tilde
m_0\frac{1}{6}c_{ijk}T_iT_jT_k+i\tilde a_0S+i\tilde c_mU_m\,
.\label{2.18}
\end{eqnarray}
However, in order to keep the algebra simple we will focus in the
following on the toroidal prepotential with $c_{ijk}=1$.

Before analyzing the  IIA situation captured by
(\ref{IIAU}) we would like to consider the simpler situation we
have advertised in Sec. \ref{2.2}, namely the
compactification without complex-structure moduli. In this case
the flux superpotential has the simple form:
\begin{equation}
W_{\mathrm{IIA}}=W_F+W_H=i\tilde e_iT_i+i\tilde
m_0T_1T_2T_3+i\tilde a_0S\, ,\label{IIA}
\end{equation}
where the last term of (\ref{IIAU}) is now missing because the
$U$'s are absent. Furthermore, the Kähler potential reads:
\begin{equation}\label{KIIA}
  K= - \log (S+\bar S)^4 \prod_{i=1}^3(T_i+\bar T_i).
\end{equation}
It is clear from the latter that this compactification is achieved
as an orbifold which effectively identifies all $U$'s and $S$. The
precise dependence of $K$ on the $S$-field is crucial for the
equations $F_S = F_{T_1}= F_{T_2}= F_{T_3}=0$ to admit a
non-trivial solution. This solution is
\begin{equation}\label{IIAsol}
|\gamma_1| T_1=|\gamma_2| T_2=|\gamma_3|T_3=
\sqrt{\frac{5|\gamma_1\gamma_2\gamma_3|}{3 \tilde m_0^2}}\, ,\quad
S=-\frac{8}{3 \tilde m_0\tilde a_0}\gamma_iT_i\, ,
\end{equation}
and describes a supersymmetric $\mathrm{AdS}_4$ vacuum. Note that
we must have $\gamma_i<0$ for this vacuum to exist.

Coming back to the more generic case with superpotential (\ref{IIAU}),
and Kähler potential $K= - \log (S+\bar S) \prod_{i=1}^3(T_i+\bar
T_i)\prod_{i=1}^3(U_i+\bar U_i)$, the equations $F_{\phi_i}=0$
admit the following solution:
\begin{equation}\label{solutionIIAg}
|\gamma_i|T_i =\sqrt{\frac{5|\gamma_1\gamma_2\gamma_3|}{3 \tilde
m_0^2}}\, ,\quad S=-\frac{2}{3 \tilde m_0\tilde a_0}\gamma_iT_i\,
,\quad \tilde c_m U_m=-\frac{2}{3 \tilde m_0}\gamma_i T_i\, .
\end{equation}
This solution corresponds again to supersymmetric $\mathrm{AdS}_4$
vacua, as shown in \cite{Camara:2005dc}.

The above analysis deserves several comments. Contrary to the case
studied in \cite{Derendinger:2004jn} where $\mathrm{AdS}_4$ with
moduli stabilized was obtained with all fluxes on, here no
geometric fluxes are present (see (\ref{IIAU})). This might seem
inconsistent at first glance. Indeed, the geometric fluxes were
introduced in Ref. \cite{Derendinger:2004jn} as a consequence of
the Jacobi identity and the tadpole conditions that the set of fluxes must satisfy. Fluxes
are indeed gauging parameters since flux compactifications are
gauged supergravities.

The resolution of this puzzle goes as follows. The background
analyzed in \cite{Derendinger:2004jn} originated from localized
branes and orientifold planes. As already mentioned in the
introduction, the vacua under investigation here {\it do not}
admit a ten-dimensional supergravity interpretation with localized
brane/orientifold sources \cite{Lust:2004ig}. One must consider
the addition of {\it smeared} brane and/or orientifold sources
\cite{acha}. This will be made more transparent in Sec.
\ref{replacing}, when discussing the brane-sources that create the
background at hand. It is clear however that in the presence of
smeared branes, Jacobi or equivalently Bianchi identities and
tadpole conditions are slightly modified and consistency is
recovered despite the non-vanishing $a_0m_0$ (see Eq.
(\ref{nafluxa})). This issue has been nicely discussed in a
similar context by Villadoro and Zwirner \cite{VZ}.

Let us end the present section by discussing some T-dual/mirror
transforms of the of the IIA models. As mentioned already,
T-duality will in general transform the NS-fluxes into geometrical
fluxes. We can for instance investigate within the toroidal models
the T-duality transformation in the internal directions $x^1$ and
$x^2$, acting as $T_1\rightarrow 1/T_1$, $T_{2,3}\rightarrow
T_{2,3}$. Then the T-dual superpotential of Eq. (\ref{IIAU})
becomes
\begin{equation}
W_{\mathrm{IIA}}=\tilde e_1 + \tilde e_2 T_1T_2+ \tilde m_0 T_2T_3
+ \tilde e_3 T_3 T_1 + \tilde a_0 S T_1 + \tilde c_mT_1U_m\,
.\label{IIAUtdual}
\end{equation}
The fluxes $\tilde a_0$ and $\tilde c_m$ become now geometrical.
The corresponding $\mathrm{AdS}_4$ ground states can be simply
obtained by replacing $T_1$ by $1/T_1$ in Eq.
(\ref{solutionIIAg}).

Alternatively let us consider the IIB mirror transform of the
superpotential (\ref{IIAU}), which is obtained by applying
T-duality transformations in the three directions $x^1$, $x^3$ and
$x^5$. This exchanges the IIA Kähler moduli by the IIB
complex-structure moduli and vice versa. In the presence of the
IIA NS-fluxes $\tilde c_m$ as in (\ref{IIAU}), the type IIB mirror
superpotential will necessarily contain geometrical fluxes:
\begin{equation}
W_{\mathrm{IIB}} = i\tilde e_i U_i + i \tilde m_0 U_1 U_2 U_3 + i
\tilde a_0 S + i \tilde c_m T_m\, .\label{IIBG}
\end{equation}
In this case, T-duality takes the system away from the original CY
framework. However, we can consider the simpler version with no
IIA complex-structure deformations and superpotential (\ref{IIA}).
Now, there are no geometrical fluxes on the mirror IIB side. This
model was indeed motivated by type IIB, where it provides the only
possibility for perturbatively stabilizing  all moduli without
geometrical fluxes. The dual geometry is not of the standard form,
since the space has no K\"ahler deformation. In fact, it is a
non-geometric space that does not allow for a large-radius
description, but rather for CFT or Landau--Ginzburg description
\cite{Becker:2007dn,Becker:2006ks}\footnote{These spaces were
discussed  by Candelas \emph{et al.} some time ago \cite{Candelas:1993nd}.}.
Ignoring the absence of a sensible large radius limit, the
mirror-dual type IIB superpotential takes the form
\begin{eqnarray}\label{tv1}
W_{\mathrm{IIB}}=\int_X\Omega\wedge
\left(F^{\mathrm{R}}_3+SH^{\mathrm{NS}}_3\right)=
ie_iU_i+im_0U_1U_2U_3+ia_0S\,  .\label{IIB}
\end{eqnarray}
Minimization of the scalar potential goes as previously explained
for type IIA (see Eqs. (\ref{IIAsol})) and this type IIB
superpotential leads to supersymmetric $\mathrm{AdS}_4$ vacua with
the $U_i$ and $S$ stabilized.

\section{The source-brane picture}
\label{replacing}

As we discussed in the previous section, supersymmetric
$\mathrm{AdS}_4$ can appear as ground states of type IIA flux
compactifications. In the simplest case, it was enough to turn on
a Neveu--Schwarz 3-form flux as well as a Ramond 4-form flux in
gauged IIA supergravity with mass parameter $\tilde m_0$ (see Eqs.
(\ref{IIAU}) and (\ref{IIA})). Generically, however, metric fluxes
are necessary. We have also stressed that Bianchi identities -- or
equivalently, Jacobi identities in the gauged-supergravity
language -- and tadpole conditions are always satisfied. This
occurs sometimes in a subtle way, in particular when
certain sources are present.

Our aim here is precisely to characterize the sources that
generate the fluxes necessary for the compactifications under
consideration. This complementary, or dual picture, gives another
perspective to the emergence of $\mathrm{AdS}_4$. The latter
appears as near-horizon geometry of a certain distribution of
intersecting/smeared branes and calibrated sources that act as domain walls,
 connecting $\mathrm{AdS}_4$ to an asymptotically
flat region. It also sheds light on the origin of the various
terms that contribute to the Bianchi identities or the tadpole
conditions. Finally, the brane picture is the first step towards
the counting of microscopic states. From the viewpoint of
four-dimensional gauged supergravity, one could presumably go
further and consider the attractor equations and the macroscopic
entropies. All this is outside the scope of the present paper.

\subsection{The brane origin of the fluxes}

As we will see explicitly in Sec. \ref{ex}, the appearance of
$\mathrm{AdS}_4$ as near-horizon geometry requires that all
branes have two common spatial directions in non-compact
four-dimensional spacetime, i.e. they have the geometry of a
domain wall in four dimensions. Moreover, depending on their
dimensionality, the branes will fill part of the internal space
$M_6$. Keeping this structure in mind, let us summarize, the
relations between the various fluxes and the corresponding source
branes.
 \begin{itemize}
 \item
 For a Neveu--Schwarz 3-form flux $H_3$ through a 3-cycle $\Sigma_3$ inside
 $M_6$ the sources are NS5-branes wrapped around the dual 3-cycle $\tilde\Sigma_3$.
 \item
 In the Ramond sector we have fluxes of the Ramond field strengths $F_n^{\mathrm{R}}$ through
 some internal $n$-dimensional cycles $\Sigma_n$. The desired domain-wall configuration in spacetime,
 requires that these fluxes be generated by magnetic brane sources, namely by D($8-n$)-branes,
 wrapped around internal cycles $\tilde\Sigma_{6-n}$
 dual to $\Sigma_n$.
 \item For  geometrical fluxes we expect to have Kaluza--Klein monopoles as sources.
 In fact, performing T-duality to directions orthogonal to the NS5-brane, one obtains a KK-monopole.
 \end{itemize}

The fluxes are quantized and this reflects that the number of
branes is not arbitrary. When $M_6$ is a CY, these numbers are
related to $h^{2,1}$ or $h^{1,1}$.

\subsection{Supersymmetric intersecting branes and calibrated sources}
\label{review}

Let us briefly review the salient features of supersymmetric
configurations of intersecting branes in supergravity, and clarify
the issues of Bianchi identities and tadpole conditions in the
presence of sources. For a comprehensive exposition of the subject
the reader is referred to the literature \cite{tsey, berg, argu,
ohta, gaun, youm, stel, boon}.

Before we proceed, it will be useful to divide the spacetime
coordinates into the following parts: (a) overall world-volume
coordinates, which are common to all branes in the system; (b)
overall transverse coordinates, which are transverse to all
branes; (c) relative transverse coordinates, which are transverse
to some (but not all) of the constituent branes. Supergravity
solutions of intersecting branes can be built according to the
``harmonic superposition rules" \cite{tsey}. The solution in this
case is constructed in terms of harmonic functions depending on
the overall transverse coordinates alone. The systems we consider
in the following will all be of this type.

In addition to the D-branes and NS5-branes of type II
supergravity, the solutions we present in the following include
supersymmetric (and therefore calibrated), smeared sources. Our
analysis is based on the following observation, which holds under
certain mild assumptions \cite{koer}: \emph{any supersymmetric
configuration with supersymmetric (and therefore calibrated), possibly smeared, sources will
automatically obey the dilaton and Einstein equations
(appropriately modified by the inclusion of the sources), provided
the source-modified Bianchi identities and form equations of
motion are also satisfied}. The analogue of this result in the
absence of sources was shown for IIA in \cite{Lust:2004ig} and for
type IIB in \cite{gaunb}. The cases of eleven-dimensional
supergravity, heterotic, type I are examined in \cite{susyM,
hetepap, typeone} respectively.

Let us consider Romans IIA supergravity \cite{roma}. The latter
includes a massive parameter $m$ which, from the point-of-view of
the ``democratic formulation" of IIA \cite{bergb}, can be thought
of as { conjugate} (but not, in general, equal) to the field
strength of a 9-form potential $C_9$. Schematically \cite{bergc}:
\begin{equation}
  \frac{\delta S_{\mathrm{bulk}}}{\delta m} = \star F_{10}\, ,
\end{equation}
where $S_{\mathrm{bulk}}$ is the action of massive IIA, and
$F_{10}:=\mathrm{d}C_9$. In all the examples we consider next, $m$
is in fact { equal} to $\star F_{10}$. This has in particular the
following consequences: (a) the supersymmetry variations of the
fermions are as in Romans supergravity, with $m$ thought of as a
field strength $F_0$ -- the Hodge-dual of $F_{10}$; (b) $m$ need
no longer be constant, i.e. $F_0$ need not be closed. The failure
of $m$ to be constant is parameterized by a one-form $j_8$ -- the
Poincar\'e dual to the world-volume of an 8-brane magnetic source:
\begin{equation}
\mathrm{d}F_0=j^{D8}~.
\end{equation}
More specifically, in the Sec. \ref{ex} we present various
supersymmetric bosonic solutions of IIA supergravity corresponding
to intersecting branes. In particular, these configurations solve
the supersymmetry equations\footnote{The transformations
(\ref{susytr}) are given in the democratic formulation; the reader
is referred to \cite{bergb,koer} for a detailed explanation of the
notation.}
\begin{align}
0 &  = \left(\nabla_\mu + \frac{1}{4} {}{H_\mu} {\mathcal{P}} +
\frac{\mathrm{e}^\Phi}{16} \sum_n {}{F_{n}} \Gamma_\mu {\mathcal{P}}_n \right) \epsilon \, , \nn\\
0 & = \left( {}{\partial} \Phi + \frac{1}{2} {}{H} {\mathcal{P}} +
\frac{\mathrm{e}^\Phi}{8} \sum_n (-1)^n (5-n) {}{F_{n}}
{\mathcal{P}}_n \right) \epsilon ~. \label{susytr}
\end{align}
In addition, all forms satisfy the Bianchi identities as well as
their equations of motion -- possibly modified by the inclusion of
{ calibrated}, smeared sources:
\begin{align}
\mathrm{d}F+H\wedge F&=-Q~j\nonumber\\
\mathrm{d}\star F-H\wedge \star F&=Q~\alpha(j)~, \label{bimod}
\end{align}
where $Q$ is the charge and $j$ is the Poincar\'e dual to the
source; the operator $\alpha$ acts by reversing the indices. In
the above we have used  polyform notation for $F$ and $j$. The
analysis of \cite{koer} then guarantees that the remaining
equations (dilaton, Einstein) are automatically
satisfied.\footnote{For the analysis of \cite{koer} to go through,
one needs to check in addition that the mixed space/time
components of the Einstein equations $E_{i0}=0$ are satisfied. In
all the examples considered in Sec. \ref{ex} of the present paper,
this is indeed the case. This readily follows from the fact that,
in each example, either (a) all fields are static, all non-zero
field-strength components are purely spatial, the metric is
diagonal, or (b) the example is T-dual to one for which (a)
holds.}

We remark that the supersymmetry transformations (\ref{susytr})
are not modified by the inclusion of sources. The latter does,
however, entail a modification of the Bianchi identities as above.
Equation (\ref{bimod}) can be accounted for by adding to the bulk
supergravity Lagrangian ($S_{\mathrm{bulk}}$) a source term schematically of the
form:
\beal S_{\mathrm{source}}=Q\int{ C\wedge \alpha(j)}-T\int{\Phi\wedge \alpha(j)} ~,
\label{ssource}
\end{align}
so that our solutions correspond to supersymmetric stationary
points of $S_{\mathrm{bulk}}+S_{\mathrm{source}}$. The first term
on the right-hand side of (\ref{ssource}) induces the modification
to the Bianchi identities. The second term
on the right-hand side accounts for the coupling of the calibrated
source to the bulk graviton, and therefore induces a modification
of the Einstein equation; $\Phi$ denotes the calibration form. In
addition, the charge ($Q$) and tension ($T$) of the source obey a
BPS condition $T=\pm Q$. The fact that supersymmetric D-branes and
orientifolds can be thought of as (generalized) calibrated sources
was shown in \cite{gencalold}.


\subsection{The tadpole equations}

Any self-consistent system of spacetime-filling branes must obey
the tadpole cancellation condition, which is a consequence of the
(generalized) Gau\ss{} law. Alternatively, in supersymmetric
configurations, this condition can be thought of as arising from
the integrated Bianchi identities (\ref{bimod}). Specifically for
the D4/D8/NS5 example of Sec. \ref{ddno}, we will need the D6/O6
tadpole cancellation condition:
\beal \frac{1}{2\pi\sqrt{\alpha^{\prime}}}\int_{\Sigma} F_0~ H_3+
N_{\mathrm{D6}}-2N_{\mathrm{O6}}=0~, \label{tadp}
\end{align}
where  $N_{\mathrm{D6}}$, $N_{\mathrm{O6}}$ is the total number
spacetime-filling D6-branes, O6-planes wrapping a three-cycle
$\Sigma$ in the internal space.

Note that from the point of view of this supergravity analysis
there is an ambiguity in the interpretation: ignoring higher-order
derivative corrections and possible world-volume excitations on
the sources, only the difference
$N_{\mathrm{D6}}-2N_{\mathrm{O6}}$ can be determined -- not the
individual numbers $N_{\mathrm{D6}}$, $N_{\mathrm{O6}}$.

\section{Specific examples}\label{ex}

We are now ready to come to our explicit examples. To facilitate
the comparison with the standard form of intersecting-brane
solutions constructed according to the harmonic superposition
rules, all solutions in the following subsections are presented in
the string frame.

Keeping in mind the above subtleties that arise in the presence of
sources, we will now consider a number of configurations of
intersecting branes and (calibrated) sources in type IIA
supergravity. They generate backgrounds which interpolate between
flat space and $\mathrm{AdS}_4$, the latter appearing as the
near-horizon geometry. The examples we consider here fall into two
classes:
\begin{itemize}
  \item Configurations corresponding to supergravity vacua with
  many flat directions ($S$ and $U_m$ fields). In all these examples, the dilaton goes to zero
  at the horizon. These include:
\begin{itemize}
  \item The double D4 distribution, where we consider two stacks
  of intersecting D4-branes. The corresponding background can
  be thought of as a vacuum of IIA supergravity with superpotential
  (\ref{IIAU}) and vanishing $\tilde{m}_0, \tilde{a}_0,
  \tilde{c}_m$: $W_{\mathrm{IIA}}=i\tilde e_iT_i$. Not all moduli are stabilized in this case.
  \item Similarly, an appropriate D2/D6 distribution generates $\mathrm{AdS}_4$, but the
  superpotential which provides this vacuum receives contributions from $F_2$ and $F_6$ (see
  Eq. (\ref{IIAF})): $W_{\mathrm{IIA}}= -\tilde{m}_i T_j T_k
  +\tilde{e}_0$. Again there are flat directions left.
  \item Performing a T-duality
  in the appropriate direction amounts to transforming $T_1$ to $1/T_1$ while leaving
  $T_{2,3}$ invariant. In this way we can map the previous configuration onto D4/D8 with
  superpotential $W_{\mathrm{IIA}}=  \tilde{e}_0 T_1 - \tilde{m}_3 T_2 -\tilde{m}_2 T_3  - \tilde{m}_1 T_1
  T_2 T_3$, i.e. (\ref{IIAU}) with $\tilde{a}_0=\tilde{c}_m = 0$.
  Again, not all moduli can be fixed.
\end{itemize}

  \item Configurations where all
  moduli are stabilized. For these
  cases it is significant that whenever stabilization is complete,
  the values of the moduli found by minimizing the scalar
  potential are recovered by a careful analysis of the spacetime
  background fields near the horizon. In particular the dilaton approaches a
  finite constant in this limit. We will show explicitly how this works
  in the example of D4/D8/NS5 distribution.

  The examples under consideration here are the following:
\begin{itemize}
  \item Configuration with D4/D8/NS5 branes. This model contains in particular
 four stacks of intersecting NS5-branes. The background is a IIA ground state
  of the superpotential (\ref{IIAU}) with \emph{all} terms non-vanishing:
  $W_{\mathrm{IIA}}=i\tilde e_iT_i+i\tilde m_0T_1T_2T_3+i\tilde a_0S+i\tilde c_mU_m$.
  This allows for full moduli stabilization (Eqs. (\ref{solutionIIAg})).
  \item The next model is obtained by performing a T-duality along one
  direction (say $x^1$). This is a type IIB model with
  D3/D5/D7/NS5/KK-branes/monoples. Its superpotential, generated by $F_1, F_3, F_5, H_3$ and geometric fluxes,
  reads $W_{\mathrm{IIB}} = i(\tilde{e}_1 U_1 + \tilde{c}_2 U_2 + \tilde{c}_3 U_3  + \tilde{c}_1 T_1 +
  \tilde{e}_2 T_2 + \tilde{e}_3 T_3 + \tilde{m}_0 U_1 T_2 T_3 + \tilde{a}_0 S)$,  and it exhibits
  an $\mathrm{AdS}_4$ vacuum with all moduli stabilized. The latter appears as the
  near-horizon geometry of the brane/monopole configuration at hand.
  A further T-duality\footnote{In total we perform a T-duality along $x^1$
  and $x^2$, which amounts to trading $T_1$ for $1/T_1$.} allows to reach a new type IIA model
  generated by intersecting D2/D6/KK-branes/monopoles with
  superpotential $W_{\mathrm{IIA}}= \tilde{e}_1 + \tilde{e}_2 T_1 T_2
  + \tilde{e}_3 T_1 T_3 + \tilde{m}_0 T_2 T_3 + \tilde{a}_0 ST_1 +
  \tilde{c}_m T_1 U_m$. This superpotential is indeed given in (\ref{IIAU}) with $F_2$ and $F_6$
  (\ref{IIAF}) and geometric fluxes  (\ref{IIAg}). All moduli are stabilized in this
  $\mathrm{AdS}_4$ mirror vacuum.
  \item Finally, we can simplify the first of the above setups by keeping only one out of four stacks of
  NS5-branes in the D4/D8/NS5. This still provides an $\mathrm{AdS}_4$ near-horizon
  geometry, but the corresponding superpotential (\ref{IIA}) has
  now vanishing $\tilde{c}_m$: $W_{\mathrm{IIA}}=i\tilde e_iT_i+i\tilde m_0T_1T_2T_3+i\tilde a_0S$.
  In this model, the IIA side was free of complex-structure moduli.
  The type IIB dual of the brane configuration is
  a D5/NS5-brane distribution which  is not the non-geometric
  type IIB vacuum free of Kähler moduli studied in Secs. \ref{2.2} and \ref{2.3}.
  The D5/NS5 has Kähler moduli and these are not stabilized.
\end{itemize}
\end{itemize}

\subsection{Examples with unstabilized moduli}

\subsubsection{D4}

Let us consider the following system of intersecting D4-branes:

\bigskip
\begin{center}
  \begin{tabular}{|c||c|c|c|c|c|c|c|c|c|c|}
    \hline
    & $\xi^0$ & $\xi^1$  & $\xi^2$  & $y^1$ & $y^2$  & $y^3$  & $x^1$  & $x^2$  & $x^3$  & $x^4$ \\
    \hline
    \hline
    $\mathrm{D}4$ & $\bigotimes$ & $\bigotimes$  & $\bigotimes$ &  &  &  & $\bigotimes$  &
               $\bigotimes$  &   &  \\

    \hline
    $\mathrm{D}4^\prime$ & $\bigotimes$ & $\bigotimes$  & $\bigotimes$ &   &   &  &   &
                 & $\bigotimes$  & $\bigotimes$ \\
\hline
   \end{tabular}
\end{center}
\bigskip
We have used the symbol $\otimes$ to denote a direction along the
brane; the boxes corresponding to directions perpendicular to the
branes have been left blank. The two D4-branes intersect on a
two-brane along the directions $\xi^\mu,~\mu=0,1,2$. The overall
transverse directions (i.e. the directions transverse to both
D4-branes) are denoted by $y^i,~i=1,2,3$, whereas the relative
transverse directions (i.e. the directions transverse to one of
the D4-branes but parallel to the other) are denoted by
$x^a,~a=1,\ldots, 4$, and may be taken to parameterize a $T^4$.
Note, however, that the D4-branes we consider here will be {\it
smeared} in both relative and  overall transverse directions.

According to the general procedure described in Sec. \ref{review},
the following can be seen to be a supersymmetric solution of IIA
supergravity in the presence of calibrated sources:
\begin{align}
\mathrm{d}s_{10}^2&=\frac{1}{\sqrt{H_1H_2}}~\eta_{\mu\nu}\mathrm{d}\xi^\mu
\mathrm{d}\xi^\nu
+\sqrt{H_1H_2}~\delta_{ij}\mathrm{d}y^i\mathrm{d}y^j +\sqrt{
\frac{H_2}{H_1}}\sum_{a=1}^2(\mathrm{d}x^a)^2
+\sqrt{\frac{H_1}{H_2}}\sum_{a=3}^4(\mathrm{d}x^a)^2
\nn\\
\mathrm{e}^{-2\phi}&=\sqrt{H_1H_2}, \quad F_{x^3x^4y^i
y^j}=-\delta^{k\ell}\varepsilon_{ijk}\partial_{y^{\ell}}H_1,\quad
F_{x^1x^2y^i
y^j}=-\delta^{k\ell}\varepsilon_{ijk}\partial_{y^{\ell}}H_2 ~,
\label{kty}
\end{align}
where ($y^2 = \delta_{ij}y^i y^j$)
\beal 4\pi
H_{\alpha}=1+\frac{c_{\alpha}}{y}+\frac{d_{\alpha}}{2y^{2}}, \quad
\alpha=1,2 \label{hadef}
\end{align}
and $c_{\alpha}$, $d_{\alpha}$ are positive constants. Moreover,
the Killing spinor is constrained to satisfy
\beal
\Gamma_{y^1y^2y^3x^1x^2}\epsilon&=-\epsilon\nn\\
\Gamma_{y^1y^2y^3x^3x^4}\epsilon&=-\epsilon \, ,
\end{align}
and therefore the system preserves one-fourth of the original
supersymmetry (eight real supercharges, or $\mathcal{N}=1$ in six
dimensions).

It is important to stress here that the functions $H_{\alpha}$ are
{\it not} harmonic in the overall transverse directions,
reflecting the presence of smeared sources. In order to assign a
physical interpretation to the above configuration note that, as
follows from (\ref{kty}),
\begin{align}
3\partial_{[y^i}F_{y^j y^k]x^1x^2}&=\varepsilon_{ijk}j_2\nn\\
3\partial_{[y^i}F_{y^j y^k]x^3x^4}&=\varepsilon_{ijk}j_1 \, ,
\end{align}
where
\beal j_{\alpha}:=-\nabla^2H_{\alpha}&= \frac{1}{4\pi
y^2}\frac{\partial}{\partial y}\left(c_{\alpha}
+\frac{d_{\alpha}}{y}\right)\nn\\
&=c_{\alpha}\delta^3(y)-\frac{d_{\alpha}}{4\pi y^4} \, ,
\label{jdef}
\end{align}
and the nabla acts in the overall transverse directions. Hence,
$j_1$ is the {\it charge density} of a four-brane magnetic source
along $\xi^0,\xi^1,\xi^2,x^1,x^2$, smeared along the overall
transverse radial direction $y$. Similarly, $j_2$ is the
 charge density of a smeared
four-brane magnetic source along $\xi^0,\xi^1,\xi^2,x^3,x^4$. The
total charge can be read off from (\ref{jdef}):
\beal Q_{\alpha} =c_{\alpha}-\frac{d_{\alpha}}{\varepsilon} ~,
\label{ambig}
\end{align}
and we have introduced a cutoff $y\geq \varepsilon$. The
configuration at hand could  be interpreted as consisting of
smeared D4-branes with ``bear" charges
$c_{\alpha}-d_{\alpha}/\varepsilon$; the divergence of the charges
could be interpreted as due to the smearing, and one could define
finite ``renormalized" charges by subtracting the infinite part.

The blowing-up of the D-brane charges and the continuous charge
distributions (\ref{jdef}) along the non-compact $y$-direction are
pathological. Moreover, it should be emphasized that smearing
along compact directions is qualitatively different from smearing
along non-compact ones: the latter would generally invalidate a
Kaluza--Klein reduction. Our point-of-view here is that this
pathological behaviour reflects the fact that the present example
corresponds to a setup with unstabilized moduli. In Sec. \ref{4.2}
we will present examples corresponding to setups with stabilized
moduli, which do not suffer from the pathologies mentioned above.

The near-horizon limit can be obtained by taking $y\rightarrow 0$.
The ten-dimensional metric takes the form
\beal
\mathrm{d}s_{\mathrm{NH}}^2=\frac{y^2}{\sqrt{D_1D_2}}~\eta_{\mu\nu}\mathrm{d}\xi^\mu
\mathrm{d}\xi^\nu
&+\sqrt{D_1D_2}~\frac{\mathrm{d}y^2}{y^2}+\sqrt{D_1D_2}~\mathrm{d}\Omega_2^2\nn\\
&+\sqrt{\frac{D_2}{D_1}}~\sum_{a=1}^2(\mathrm{d}x^a)^2
+\sqrt{\frac{D_1}{D_2}}~\sum_{a=3}^4(\mathrm{d}x^a)^2~, \label{nh}
\end{align}
where $\mathrm{d}\Omega_2^2$ is the metric of the unit two-sphere
in the overall transverse directions, and we have set
$D_{\alpha}:=d_{\alpha}/8\pi$. We can readily recognize (\ref{nh})
as the metric of $\mathrm{AdS}_4\times S^2\times T^4$. At
$y\rightarrow\infty$ the metric asymptotes $\mathbb{R}^{1,5}\times
T^4$. Finally note that, as follows from (\ref{kty}, \ref{hadef}),
the string coupling $g_\mathrm{s}=\mathrm{e}^{\phi}$ goes to zero
as $y\rightarrow 0$. This is a typical runaway behaviour. In order
to stabilize the string coupling to a finite value one has to add
NS5-branes, as we will see in Sec. \ref{4.2}.

\subsubsection{D2/D6}

This configuration can be obtained from the previous one by
applying two T-duality transformations along directions $x^1,x^2$.
More
explicitly, the branes are as follows:

\bigskip
\begin{center}
  \begin{tabular}{|c||c|c|c|c|c|c|c|c|c|c|}
    \hline
    & $\xi^0$ & $\xi^1$  & $\xi^2$  & $y^1$ & $y^2$  & $y^3$  & $x^1$  & $x^2$  & $x^3$  & $x^4$ \\
    \hline
    \hline
    $\mathrm{D}2$ & $\bigotimes$ & $\bigotimes$  & $\bigotimes$ &  &  &  &  &
                &   &  \\

    \hline
    $\mathrm{D}6$ & $\bigotimes$ & $\bigotimes$  & $\bigotimes$ &   &   &  &  $\bigotimes$  &  $\bigotimes$
                 & $\bigotimes$  & $\bigotimes$ \\
\hline
   \end{tabular}
\end{center}
\bigskip
The metric now reads:
\begin{align}
\mathrm{d}s_{10}^2&=\frac{1}{\sqrt{H_1H_2}}~\eta_{\mu\nu}\mathrm{d}\xi^\mu
\mathrm{d}\xi^\nu
+\sqrt{H_1H_2}~\delta_{ij}\mathrm{d}y^i\mathrm{d}y^j
+\sqrt{\frac{H_1}{H_2}}\sum_{a=1}^4(\mathrm{d}x^a)^2~,
\end{align}
with $H_{\alpha}$ as in (\ref{hadef}). As in the previous
subsection, in the near-horizon limit the metric approaches
$\mathrm{AdS}_4\times S^2\times T^4$ and the string coupling goes
to zero.

\subsubsection{D4/D8}

Extra T-duality along $y^2, y^3$ leads to the following D4/D8
brane distribution:

\bigskip
\begin{center}
  \begin{tabular}{|c||c|c|c|c|c|c|c|c|c|c|}
    \hline
    & $\xi^0$ & $\xi^1$  & $\xi^2$  & $y^1$ & $y^2$  & $y^3$  & $x^1$  & $x^2$  & $x^3$  & $x^4$ \\
    \hline
    \hline
    $\mathrm{D}4$ & $\bigotimes$ & $\bigotimes$  & $\bigotimes$ &  & $\bigotimes$ & $\bigotimes$ &  &
                &   &  \\

    \hline
    $\mathrm{D}8$ & $\bigotimes$ & $\bigotimes$  & $\bigotimes$ &   &  $\bigotimes$ & $\bigotimes$ &  $\bigotimes$  &  $\bigotimes$
                 & $\bigotimes$  & $\bigotimes$ \\
\hline
   \end{tabular}
\end{center}
\bigskip
The discussion of this model is analogous to the previous two
cases.

\subsection{Examples with all moduli stabilized}\label{4.2}

\subsubsection{D4/D8/NS5}\label{ddno}

We now come to the domain-wall solution of most interest, since it
may be thought of as the brane-dual to the supersymmetric
$\mathrm{AdS}_4$ vacuum considered in Sec. \ref{2.3}. Consider the
following system of intersecting D4/D8/NS5-branes:
\bigskip
\begin{center}
  \begin{tabular}{|c||c|c|c|c|c|c|c|c|c|c|}
    \hline
    & $\xi^0$ & $\xi^1$  & $\xi^2$  & $y$ & $x^1$  & $x^2$  & $x^3$  & $x^4$  & $x^5$  & $x^6$ \\
    \hline
    \hline
    $\mathrm{D}4$ & $\bigotimes$ & $\bigotimes$  & $\bigotimes$ &  &$\bigotimes$  &  $\bigotimes$ &   &
                &   &  \\

    \hline
    $\mathrm{D}4^{\prime}$ & $\bigotimes$ & $\bigotimes$  & $\bigotimes$ &  &  &  & $\bigotimes$  &
               $\bigotimes$  &   &  \\

    \hline
    $\mathrm{D}4^{\prime\prime}$ & $\bigotimes$ & $\bigotimes$  & $\bigotimes$ &   &   &  &   &
                 & $\bigotimes$  & $\bigotimes$ \\
\hline
    $\mathrm{NS}5$ & $\bigotimes$ & $\bigotimes$  & $\bigotimes$ &   &  $\bigotimes$  &   &  $\bigotimes$ &
                 & $\bigotimes$  & \\
\hline
    $\mathrm{NS}5^{\prime}$ & $\bigotimes$ & $\bigotimes$  & $\bigotimes$ &   &  $\bigotimes$  &  &   &  $\bigotimes$
                 &  & $\bigotimes$ \\
 \hline
    $\mathrm{NS}5^{\prime\prime}$ & $\bigotimes$ & $\bigotimes$  & $\bigotimes$ &   &  & $\bigotimes$ &   &
               $\bigotimes$    &  $\bigotimes$  & \\
\hline
    $\mathrm{NS}5^{\prime\prime\prime}$ & $\bigotimes$ & $\bigotimes$  & $\bigotimes$ &   &  & $\bigotimes$ &  $\bigotimes$     &
               &  &  $\bigotimes$ \\
 \hline
    $\mathrm{D}8$ & $\bigotimes$ & $\bigotimes$  & $\bigotimes$ &   &  $\bigotimes$ & $\bigotimes$ & $\bigotimes$  &  $\bigotimes$
                 & $\bigotimes$  & $\bigotimes$ \\
\hline
   \end{tabular}
\end{center}
\bigskip
These generate a supersymmetric solution of IIA supergravity in
the presence of supersymmetric (calibrated) sources. As we will
see shortly, the tadpole cancellation condition induces in
addition O6-planes and/or D6-branes. These brane distributions
should thus be referred to as D4/D8/NS5/O6/D6, which we will avoid
for simplicity, keeping only D4/D8/NS5.

\paragraph{Background fields and tadpole cancellation.}

The explicit form of the solution is given by:
\begin{center}
\fbox{\parbox{12.6cm}{
\begin{align}
\mathrm{d}s_{10}^2=&\left\{{H^{\mathrm{D}8}\left(\prod_{\alpha=1}^3
H^{\mathrm{D}4}_{\alpha} \right)
 } \right\}^{-\frac{1}{2}}~
\eta_{\mu\nu}\mathrm{d}\xi^\mu \mathrm{d}\xi^\nu\nn\\
&+\left( \prod_{\alpha=1}^4 H^{\mathrm{NS}5}_{\alpha}\right)
\left\{{H^{\mathrm{D}8}\left(\prod_{\alpha=1}^3
H^{\mathrm{D}4}_{\alpha} \right)
  } \right\}^{\frac{1}{2}}~\mathrm{d}y^2\nn\\
&+\sqrt{\frac{H_{2}^{\mathrm{D}4}H_{3}^{\mathrm{D}4}}{H_{1}^{\mathrm{D}4}H^{\mathrm{D}8}_{}}}\left\{
H_{3}^{\mathrm{NS}5}H_{4}^{\mathrm{NS}5}~(\mathrm{d}x^1)^2
+H_{1}^{\mathrm{NS}5}H_{2}^{\mathrm{NS}5}~(\mathrm{d}x^2)^2\right\}\nn\\
&+\sqrt{\frac{H_{1}^{\mathrm{D}4}H_{3}^{\mathrm{D}4}}{H_{2}^{\mathrm{D}4}H^{\mathrm{D}8}_{}}}\left\{
H_{2}^{\mathrm{NS}5}H_{3}^{\mathrm{NS}5}~(\mathrm{d}x^3)^2
+H_{1}^{\mathrm{NS}5}H_{4}^{\mathrm{NS}5}~(\mathrm{d}x^4)^2 \right\}\nn\\
&+\sqrt{\frac{H_{1}^{\mathrm{D}4}H_{2}^{\mathrm{D}4}}{H_{3}^{\mathrm{D}4}H^{\mathrm{D}8}_{}}}\left\{
H_{2}^{\mathrm{NS}5}H_{4}^{\mathrm{NS}5}~(\mathrm{d}x^5)^2
+H_{1}^{\mathrm{NS}5}H_{3}^{\mathrm{NS}5}~(\mathrm{d}x^6)^2\right\};\nn\\
\mathrm{e}^{2\phi}&= \left( \prod_{\alpha=1}^4
H^{\mathrm{NS}5}_{\alpha}\right)^{}
\left(\prod_{\alpha=1}^3 H^{\mathrm{D}4}_{\alpha} \right) ^{-\frac{1}{2}}    \left(H^{\mathrm{D}8}\right)^{-\frac{5}{2}}    ;\nn\\
H_{x^2x^4x^6}&=-\partial_y
H^{\mathrm{NS}5}_1\left(H^{\mathrm{D}8}\right)^{-1};
~~~H_{x^2x^3x^5}=-\partial_y H^{\mathrm{NS}5}_2\left(H^{\mathrm{D}8}\right)^{-1};\nn\\
H_{x^1x^3x^6}&=-\partial_y
H^{\mathrm{NS}5}_3\left(H^{\mathrm{D}8}\right)^{-1};
~~~H_{x^1x^4x^5}=-\partial_y H^{\mathrm{NS}5}_4\left(H^{\mathrm{D}8}\right)^{-1}; \nn\\
F_{x^3x^4x^5x^6}&=\partial_{y}H^{\mathrm{D}4}_1; ~~~ F_{x^1x^2x^5x^6}=\partial_{y}H^{\mathrm{D}4}_2;\nn\\
F_{x^1x^2x^3x^4}&=\partial_{y}H^{\mathrm{D}4}_3; ~~~F=-\partial_y
H^{\mathrm{D}8}\left( \prod_{\alpha=1}^4
H^{\mathrm{NS}5}_{\alpha}\right)^{-1} ~.\nn
\end{align}
}}
\end{center}
\beal\label{ktyb}\end{align}

The ten-dimensional Killing spinor is constrained to satisfy
\beal
\Gamma_y\epsilon&=\epsilon\nn\\
\Gamma_{x^1x^2x^3x^4}\epsilon&=-\epsilon\nn\\
\Gamma_{x^3x^4x^5x^6}\epsilon&=-\epsilon\nn\\
\Gamma_{x^2x^4x^6}\epsilon&=-\Gamma_{11}\epsilon ~,
\label{calmo}
\end{align}
therefore the system preserves one-sixteenth of the original
supersymmetry (two real supercharges, or $\mathcal{N}=1$ in three
dimensions). More details can be found in appendix \ref{appsus}.

It can be explicitly checked that (\ref{ktyb}) solves the
supersymmetry equations as well as the
equations-of-motion for the forms. In addition, the fluxes obey source-modified
Bianchi identities. It then follows from the integrability theorem
of \cite{koer} that the correspondingly source-modified dilaton
and Einstein equations are automatically obeyed\footnote{ Strictly
speaking this statement was only proven in \cite{koer} for
D-branes and orientifold planes; we expect that a generalization
to include NS5-brane sources should be straightforward.}. In other
words, (\ref{ktyb}) is a {\it family} of solutions of type IIA
supergravity, parameterized by the functions
$H_{\alpha}^{\mathrm{D}4}$, $H^{\mathrm{D}8}$,
$H_{\alpha}^{\mathrm{NS}5}$. The modification of the Bianchi
identities due the presence of the sources is given by
\begin{align}
\partial_{y}F_{4}&=j^{\mathrm{D}4}\nn\\
\partial_{y}F_{0}&=j^{\mathrm{D}8}\nn\\
\partial_{y}H_3&=j^{\mathrm{NS}5}
~,\label{biviol}
\end{align}
where the charge densities $j$ are read off from (\ref{ktyb}) and
the equations above. The sourceless case corresponds to setting
the right-hand side to zero, $j=0$.

We observe from (\ref{ktyb}) that in the ``massless" IIA case
(corresponding to $F_0=0$, $H^{\mathrm{D}8}=\mathrm{const.}$) the
$F_4$, $H_3$ fluxes satisfy their corresponding sourceless Bianchi
identities if and only if the functions
$H^{\mathrm{D}4}_{\alpha}$, $H_{\alpha}^{\mathrm{NS}5}$ are
harmonic in the total transverse space (i.e. the $y$-direction).
This is no longer true in the massive case, $F_0\neq 0$. Note also
that the solution has $F_2=0$. Consistency with the tadpole
cancellation condition (\ref{tadp}) then induces the presence of a
non-zero charge density corresponding to O6-planes and/or
D6-branes:\footnote{It can be seen that the net charge of the
configuration is that of orientifold six-planes. This is in
agreement with the analysis of \cite{acha}, which becomes relevant
in the near-horizon limit as we will see in the following.}
\beal j^{\mathrm{O6/D6}}_{\alpha}=\partial_yH^{\mathrm{D8}}~
\partial_yH^{\mathrm{NS5}}_{\alpha}\left(H^{\mathrm{D}8} \prod_{\beta=1}^4 H^{\mathrm{NS}5}_{\beta}\right)^{-1}~,
\end{align}
where we have used (\ref{ktyb}) to express $F_0$ and $H_3$ in
terms of $H^{\mathrm{D8}}$, $H^{\mathrm{NS5}}_{\alpha}$,
respectively.
The induced  O6-planes and/or D6-branes are along
$(\xi^\mu,y,x^{1,3,5})$, $(\xi^\mu,y,x^{1,4,6})$,
$(\xi^\mu,y,x^{2,4,5})$ and $(\xi^\mu,y,x^{2,3,6})$:
\bigskip
\begin{center}
  \begin{tabular}{|c||c|c|c|c|c|c|c|c|c|c|}
    \hline
    & $\xi^0$ & $\xi^1$  & $\xi^2$  & $y$ & $x^1$  & $x^2$  & $x^3$  & $x^4$  & $x^5$  & $x^6$ \\
    \hline
    \hline
    $\mathrm{O}6$ & $\bigotimes$ & $\bigotimes$  & $\bigotimes$ & $\bigotimes$  &$\bigotimes$  &    &$\bigotimes$ &
               &    $\bigotimes$ &  \\

    \hline
    $\mathrm{O}6^{\prime}$ & $\bigotimes$ & $\bigotimes$  & $\bigotimes$ &   $\bigotimes$ & $\bigotimes$ &  & &
               $\bigotimes$  &   &    $\bigotimes$\\

    \hline
    $\mathrm{O}6^{\prime\prime}$ & $\bigotimes$ & $\bigotimes$  & $\bigotimes$ &    $\bigotimes$ &     &  $\bigotimes$ &   &
                 $\bigotimes$  & $\bigotimes$  &  \\
\hline
    $\mathrm{O}6^{\prime\prime\prime}$ & $\bigotimes$ & $\bigotimes$  & $\bigotimes$ &   $\bigotimes$  &   &
 $\bigotimes$ &  $\bigotimes$ &
                 &   &   $\bigotimes$\\
\hline
   \end{tabular}\label{tabela}
\end{center}
\bigskip
This is
consistent with the supersymmetric intersection rule
$\mathrm{NS}5\cap \mathrm{D}p(p-3)$ as well as $\mathrm{NS}5\cap
\mathrm{D}p(p-1)$. Furthermore, the 0- and 2-brane tadpole
cancellation conditions are automatically satisfied since
$H_3\wedge F_4$, $H_3\wedge F_6$ vanish, as can be checked
explicitly.

Associated with each O6 orientifold there is an internal-space
involution $\sigma_{\alpha}$, $\alpha\in \{1,\ldots, 4\}$,
reversing the parity in the directions orthogonal to the
six-plane. Explicitly, $\sigma_1: (x^2,x^4,x^6)\rightarrow
(-x^2,-x^4,-x^6)$, etc. Under each of the four orientifold
involutions, the background fields $\phi$, $g$, $F_{0}$, $F_{4}$
are even, while $H_3$, $F_{2}$ are odd. It can readily be checked
that this is indeed the case for the solution (\ref{ktyb}). In
addition, it can be checked that the involutions leave the
Killing-spinor conditions (\ref{calmo}) invariant.

The total charges can be read off from (\ref{ktyb}):
\beal Q^{\mathrm{D}4}&:=\int{\mathrm{d}y}~
j^{\mathrm{D}4}
\nn\\
Q^{\mathrm{D}8}&:=\int{\mathrm{d}y}~
j^{\mathrm{D}8}
\nn\\
Q^{\mathrm{NS}5}&:=\int{\mathrm{d}y}~
j^{\mathrm{NS}5} ~.\label{chara}
\end{align}
Similarly the
orientifold/D6-brane charge,
$Q^{\mathrm{O}6/\mathrm{D}6}_{\alpha}$, is given by the integral over the
three-cycle $\Sigma_{\alpha}\in T^6$ which the orientifold
plane/D6-brane is wrapping:
\begin{equation}
Q^{\mathrm{O}6/\mathrm{D}6}_\alpha:=\int_{\Sigma}~
\mathrm{dVol}~j^{\mathrm{O}6/\mathrm{D}6}_\alpha~.\label{charb}
\end{equation}
Note, however, that the resulting charge may in general have a
$y$-dependence. For appropriate choices of $H^{\mathrm{D8}}$,
$H^{\mathrm{NS5}}_{\alpha}$ it can be arranged that there is no
$y$-dependence. This will indeed be the case in the near-horizon
limit which we examine below.

\paragraph{The near-horizon limit.}

As already stressed, (\ref{ktyb}) corresponds to a family of
solutions. There are special choices of the $H$-functions which
lead to an $\mathrm{AdS}_4\times T^6$ near-horizon limit. Consider
for example the following:
\beal H^{\mathrm{NS}}_{\alpha}&=\left\{
\begin{array}{ll}
{c^{\mathrm{NS}}_{\alpha}}{{y}}\left\{
1+\frac{3}{2}\left(\frac{{y}}{y_0}\right)^{-\frac{5}{3}}\right\},\quad  & {y}<y_0 \\
\frac{5}{2}{c^{\mathrm{NS}}_{\alpha}}{y_0},\quad & {y}\geq y_0
\end{array}\right.\nn\\
H^{\mathrm{D}4}_{\alpha}&=\left\{
\begin{array}{ll}
{c^{\mathrm{D}4}_{\alpha}}{{y}}\left\{
1-\frac{1}{2}\left(\frac{{y}}{y_0}\right)^{}\right\},\; \, \qquad  & {y}<y_0 \\
\frac{1}{2}{c^{\mathrm{D}4}_{\alpha}}{y_0},\; \,\qquad   & {y}\geq
y_0
\end{array}\right.
 \nn\\
H^{\mathrm{D}8}&=\left\{
\begin{array}{ll}
c^{\mathrm{D}8}_{}{{y}}\left\{
1+\frac{3}{5}\left(\frac{ y}{y_0}\right)^{-\frac{8}{3}}\right\},\quad  & {y}<y_0 \\
\frac{8}{5}c^{\mathrm{D}8}_{}{y_0}, \quad & {y}\geq y_0\, ,
\end{array}\right.
\label{hbdef}
\end{align}
where the $c_{\alpha}$ are constants ($\alpha \in \{1,2,3\}$ for
D4 and $\{1,2,3,4\}$ for NS5).
The leading $y$-dependence of the $H$'s in the ``near-horizon"
$y\rightarrow 0$ limit is \emph{uniquely} determined by the
requirement that the dilaton should approach a finite constant at
the horizon. The subleading behaviour is, however, largely
arbitrary. We have partially constrained it by imposing  that $H$
should contain a harmonic piece, the piece linear in $y$, to make
contact with the corresponding solution in the absence of smeared
sources; additionally we have required that $\partial_y H$ should
be continuous. Moreover, we have introduced an (arbitrary) point
$y_0$ on the $y$-axis and imposed that $\partial_y
H^{\mathrm{D8}}$, $\partial_y H^{\mathrm{D4}}$, $\partial_y
H^{\mathrm{NS}5}$ vanish for $y\geq y_0$; this ensures that for
$y\geq y_0$ the solution is that of flat space.

The requirements of the previous paragraph do not completely
determine the $H$-functions, and we have made the particular
choices in (\ref{hbdef}) for the sake of concreteness. An
alternative choice with the exact same near-horizon limit as
(\ref{hbdef}), but which ensures that $Q_{\alpha}^{\mathrm{O6/D6}
}$ of (\ref{charb}) is everywhere
$y$-independent\footnote{Equation (\ref{hbdef}) leads to O6/D6
charges which are $y$-independent only in the near-horizon
limit.}, is as follows. Define $H^{D8}$ by
\beal
H^{\mathrm{D}8}&=\left\{
\begin{array}{ll}
c^{\mathrm{D}8} \mathrm{exp}\left\{
\frac{10}{9}\frac{d^{\mathrm{NS}5}_1}{\Pi_{\alpha=1}^4
d^{\mathrm{NS}5}_{\alpha}} \int \mathrm{d}y
\frac{\Pi_{\alpha=1}^4 H^{\mathrm{NS}5}_{\alpha}}{\partial_y
H^{\mathrm{NS}5}_1} \right\}
,\quad  & {y}\leq y_0 \\
H^{\mathrm{D}8}(y_0), \quad & {y}> y_0\, ,
\end{array}\right.
\label{4.16}
\end{align}
with all other $H$'s as in (\ref{hbdef}), and
$d^{\mathrm{NS}5}:=3/2~y_0^{5/3}c^{\mathrm{NS}5}$. In the
near-horizon limit we have $H^{\mathrm{D}8}\sim y^{-5/3}$ as
before; moreover it follows that $F_{0}H_{3}$ is constant, leading
to $y$-independent O6/D6 charges. The difference with
(\ref{hbdef}) is that $\partial_y H^{\mathrm{D8}}$ is now
discontinuous at $y=y_0$, signaling the presence of an eight-brane
source at this point on the $y$-axis. Note also that $j^{O6/D6}=0$
for $y>y_0$, which implies that the O6 planes end on the
eight-brane located at $y=y_0$.

We emphasize that, as can be seen from (\ref{hbdef}), (\ref{4.16}), all $H$'s and $\partial_y H$'s are continuous, except for $\partial_y H^{D8}$. The later has a finite jump at $y=y_0$, corresponding to a finite jump in
the `Roman's mass', as should be expected of domain wall solutions patching asymptotic regions
with different `cosmological constants'.

Anticipating the $\mathrm{AdS}_4$ near-horizon limit, we are
considering the solution only on the semi-infinite line $y>0$. One
could trivially ``glue" a copy of the solution on the $y<0$ side
by replacing $H(y)\rightarrow H(|y|)$. This would of course result
in a discontinuity of the flux at $y=0$, which could be attributed
to the presence of a brane at the origin.

We now come to the crucial point of recovering the supersymmetric
$\mathrm{AdS}_4$ as the near-horizon limit of the metric, reached
at $y\to 0$. The ten-dimensional metric takes the form
\beal
\mathrm{d}s_{\mathrm{NH}}^2&=\left\{{d^{\mathrm{D}8}\left(\prod_{\alpha=1}^3
d^{\mathrm{D}4}_{\alpha} \right)
 } \right\}^{-\frac{1}{2}}~{y}^{-\frac{2}{3}}
\eta_{\mu\nu}\mathrm{d}\xi^\mu \mathrm{d}\xi^\nu +\left(
\prod_{\alpha=1}^4 d^{\mathrm{NS}5}_{\alpha}\right)
\left\{{d^{\mathrm{D}8}\left(\prod_{\alpha=1}^3
d^{\mathrm{D}4}_{\alpha} \right)
  } \right\}^{\frac{1}{2}}~\frac{\mathrm{d}y^2}{y^2}\nn\\
&+\sqrt{\frac{d_{2}^{\mathrm{D}4}d_{3}^{\mathrm{D}4}}{d_{1}^{\mathrm{D}4}d^{\mathrm{D}8}_{}}}\left\{
d_{3}^{\mathrm{NS}5}d_{4}^{\mathrm{NS}5}~(\mathrm{d}x^1)^2
+d_{1}^{\mathrm{NS}5}d_{2}^{\mathrm{NS}5}~(\mathrm{d}x^2)^2\right\}\nn\\
&+\sqrt{\frac{d_{1}^{\mathrm{D}4}d_{3}^{\mathrm{D}4}}{d_{2}^{\mathrm{D}4}d^{\mathrm{D}8}_{}}}\left\{
d_{2}^{\mathrm{NS}5}d_{3}^{\mathrm{NS}5}~(\mathrm{d}x^3)^2
+d_{1}^{\mathrm{NS}5}d_{4}^{\mathrm{NS}5}~(\mathrm{d}x^4)^2 \right\}\nn\\
&+\sqrt{\frac{d_{1}^{\mathrm{D}4}d_{2}^{\mathrm{D}4}}{d_{3}^{\mathrm{D}4}d^{\mathrm{D}8}_{}}}\left\{
d_{2}^{\mathrm{NS}5}d_{4}^{\mathrm{NS}5}~(\mathrm{d}x^5)^2
+d_{1}^{\mathrm{NS}5}d_{3}^{\mathrm{NS}5}~(\mathrm{d}x^6)^2\right\}
~, \label{nbh}
\end{align}
where we have defined
$d^{\mathrm{D}4}:=c^{\mathrm{D}4}$ and
$d^{\mathrm{D}8}:=3/5~y_0^{8/3}c^{\mathrm{D}8}$, in addition to
$d^{\mathrm{NS}5}:=3/2~y_0^{5/3}c^{\mathrm{NS}5}$ defined previously.
We can readily
recognize (\ref{nbh}) as the metric of $\mathrm{AdS}_4\times T^6$.
 At
$y\rightarrow\infty$ the metric asymptotes $\mathbb{R}^{1,3}\times
T^6$. Note also that, as follows from (\ref{ktyb},
\ref{hbdef}), the string coupling $g_\mathrm{s}=
\mathrm{e}^{\phi}$ approaches a finite constant in the
$y\rightarrow 0$ limit:
\begin{equation}
  \mathrm{e}^{\phi}\longrightarrow \left( \prod_{\alpha=1}^4
d^{\mathrm{NS}5}_{\alpha}\right)^{\frac{1}{2}}
\left(\prod_{\alpha=1}^3 d^{\mathrm{D}4}_{\alpha}
\right)^{-\frac{1}{4}} \left(d^{\mathrm{D}8}\right)^{-\frac{5}{4}}
~.
\end{equation}
Moreover, all the flux components take on constant values:
\beal
H_{x^2x^4x^6}&=\frac{2}{3}d^{\mathrm{NS}5}_1\left(d^{\mathrm{D}8}\right)^{-1};
~~~H_{x^2x^3x^5}=\frac{2}{3}d^{\mathrm{NS}5}_2\left(d^{\mathrm{D}8}\right)^{-1} \nn\\
H_{x^1x^3x^6}&=\frac{2}{3}d^{\mathrm{NS}5}_3\left(d^{\mathrm{D}8}\right)^{-1};
~~~H_{x^1x^4x^5}=\frac{2}{3}d^{\mathrm{NS}5}_4\left(d^{\mathrm{D}8}\right)^{-1}\nn\\
F_{x^3x^4x^5x^6}&=d^{\mathrm{D}4}_1; ~~~ F_{x^1x^2x^5x^6}=d^{\mathrm{D}4}_2;~~~
F_{x^1x^2x^3x^4}=d^{\mathrm{D}4}_3; ~~~F=\frac{5}{3}d^{\mathrm{D}8}\left( \prod_{\alpha=1}^4
d^{\mathrm{NS}5}_{\alpha}\right)^{-1} ~.\label{rtyu}
\end{align}
As we shall see shortly (cf. Eqs. (\ref{iot}, \ref{chrg}) below),
this implies that all charges are regular in the near-horizon
limit. In addition, from (\ref{biviol}, \ref{rtyu}) we see that
the Poincar\'e duals to the D4/D8/NS5 all vanish in the limit,
while the Poincar\'e duals to the orientifold six-planes/D6-branes
become constant. In other words, \emph{zooming in on the
near-horizon region amounts to replacing the D4/D8/NS5 branes with
fluxes, while adding a set of completely smeared orientifold
six-planes}. This is a perfectly consistent picture from the
ten-dimensional supergravity point-of-view, as we will now show.
Our discussion will also establish that at the horizon there is a
supersymmetry enhancement from $\mathcal{N}=1$ (two real
supercharges) in three dimensions to $\mathcal{N}=1$ in four
dimensions (four real supercharges).

\paragraph{An important remark.}

The most general form of $\mathcal{N}=1$ $\mathrm{AdS}_4$ compactifications of IIA
supergravity on manifolds of $SU(3)$-structure
 was given by two of the present authors in
\cite{Lust:2004ig}. Since the group structure of $T^6$ is trivial,
it is certainly contained in $SU(3)$; it is therefore natural to
ask whether we can recover the near-horizon solution given above,
as a limiting case of the solutions of \cite{Lust:2004ig}. We will
now show that this is indeed the case.

Setting the dilaton to zero, the solutions
of \cite{Lust:2004ig} are given by (the reader is referred to the original paper for further details):
\beal\label{ltsol}
F_2&=\frac{f}{18}J+\tilde{F}\nn\\
H&=\frac{4m}{5}\mathrm{Re}(\Omega)\nn\\
F_4&=f\mathrm{dVol}_4+\frac{3m}{5} J\wedge J\nn\\
W&=-\frac{1}{5}m+\frac{i}{6}f ~.
\end{align}
In the above ($J$, $\Omega$) is the $SU(3)$ structure of the
internal six-manifold and $f$, $m$ are constants parameterizing
the solution: $f$ is the Freund--Rubin parameter, while $m$ is the
mass of Romans' supergravity -- which can be identified with $F_0$
in the ``democratic" formulation. The curvature of the
$\mathrm{AdS}_4$ is proportional to $|W|^2$. The two-form
$\tilde{F}$ is the primitive part of $F_2$ (i.e. it is in the
$\bf{8}$ of $SU(3)$) and is constrained by the Bianchi identity:
\beal \mathrm{d}\tilde{F}=\frac{1}{27}\left(\frac{108}{5}m^2-f^2
\right)\mathrm{Re}(\Omega)+j^{D6/O6} ~,\label{ltsolb}
\end{align}
where we have added a source for D6-branes/O6-planes on the
right-hand side\footnote{ The addition of this source term was
first considered in \cite{acha}.}. Finally, the only non-zero
torsion classes of the internal manifold are given by:
\beal
{\cal W}^-_1&=-\frac{4i}{9}f; ~~~~~
{\cal W}^-_2=-2i\tilde{F} ~.
\end{align}
We observe that the near-horizon limit of the intersecting-brane
solution given in the previous paragraph corresponds to setting
$\tilde{F}=0$ and taking the $f\rightarrow 0$ limit of
(\ref{ltsol}, \ref{ltsolb}), while adding the following source
term:
\beal j^{D6/O6}=- \frac{4m^2}{5} \mathrm{Re}(\Omega)~.
\end{align}
Indeed, in this limit all torsion classes of the internal manifold vanish (as they should for $T^6$),
while the flux content exactly corresponds to (\ref{rtyu}).

The $\tilde{F}=0$, $f=0$ limit and its phenomenological
implications  were considered in detail by Acharya \emph{et al.}
in \cite{acha}.

\paragraph{Comparison with the effective-superpotential approach.}

The configuration (\ref{ktyb}, \ref{hbdef}) could be interpreted
as consisting of (smeared) intersecting D4/D8/NS5-branes with
charges $Q^{\mathrm{D4}}$, $Q^{\mathrm{D8}}$, $Q^{\mathrm{NS}5}$,
together with orientifold O6-planes/D6-branes. In order to make
contact with the effective superpotential treatment of Sec.
\ref{2.3}, Eq. (\ref{IIAU}), one should use the following
dictionary:
\begin{align}\label{iot}
Q_i^{\mathrm{D}4}&= \tilde e_i\nn\\
Q^{\mathrm{D}8}&= \tilde m_0 ~\nn\\
Q^{\mathrm{NS}5}_1&=\tilde a_0;~~~~~ Q^{\mathrm{NS}5}_{i+1}=\tilde c_i, ~~i=1,2,3~,
\end{align}
where we recognize the contributions of $F_4$, $F_0$ and $H_3$.
The tadpole equation (\ref{nafluxa}) is then reproduced for
$Q^{\mathrm{O}6/\mathrm{D}6}_\alpha=
\tilde{N}^{\mathrm{flux}}_\alpha$. Moreover, substituting the
near-horizon limit of (\ref{hbdef}) in (\ref{chara}, \ref{biviol})
and taking (\ref{iot}) into account, we obtain
\beal\label{chrg}
\tilde e_i&=c_i^{\mathrm{D4}}\nn\\
\tilde m_0&= c^{\mathrm{D8}}\left(\Pi_{\alpha=1}^4 c_{\alpha}^{\mathrm{NS5}} \right)^{-1} ~\nn\\
\tilde a_0 &=c^{\mathrm{NS}5}_1\left(c^{\mathrm{D8}} \right)^{-1};~~~~~
\tilde c_i=c^{\mathrm{NS}5}_{i+1}\left(c^{\mathrm{D8}} \right)^{-1}, ~~i=1,2,3~.
\end{align}
For simplicity, here and in the remainder we ignore
numerical normalizations and set $y_0=1$.

We would like to recover the stabilized values for the moduli,
obtained in the effective-superpotential description, from the
near-horizon limit of the supergravity solution discussed above.
First, we would like to identify the geometric moduli
corresponding to solution (\ref{ktyb}) with near-horizon geometry
(\ref{nbh}-\ref{rtyu}). The internal $T^6$ can be thought of as the product
of three two-tori corresponding to $(x^1,x^2)$, $(x^3,x^4)$ and
$(x^5,x^6)$ -- in accordance with the
orientifold projections. There are then three K\"{a}hler moduli $T_i$,
corresponding  to the areas of the three $T^2$'s. Explicitly:
\beal\label{kahlm}
T_1&=\sqrt{\frac{c_2^{\mathrm{D4}}c_3^{\mathrm{D4}}}{c_1^{\mathrm{D4}}c^{\mathrm{D8}}}}\sqrt{\Pi_{\alpha=1}^4 c_{\alpha}^{\mathrm{NS5}} }\nn\\
T_2&=\sqrt{\frac{c_1^{\mathrm{D4}}c_3^{\mathrm{D4}}}{c_2^{\mathrm{D4}}c^{\mathrm{D8}}}}\sqrt{\Pi_{\alpha=1}^4 c_{\alpha}^{\mathrm{NS5}} }\nn\\
T_3&=\sqrt{\frac{c_1^{\mathrm{D4}}c_2^{\mathrm{D4}}}{c_3^{\mathrm{D4}}c^{\mathrm{D8}}}}\sqrt{\Pi_{\alpha=1}^4 c_{\alpha}^{\mathrm{NS5}} }~.
\end{align}
Moreover there are three complex structures $\tau_i$,
corresponding to the ratios of the radii of each of the $T^2$'s.
Explicitly:
\beal
\tau_1=\sqrt{\frac{c_1^{\mathrm{NS5}}c_2^{\mathrm{NS5}}}{c_3^{\mathrm{NS5}}c_4^{\mathrm{NS5}}}}; ~~~~~
\tau_2=\sqrt{\frac{c_1^{\mathrm{NS5}}c_4^{\mathrm{NS5}}}{c_2^{\mathrm{NS5}}c_3^{\mathrm{NS5}}}}; ~~~~~
\tau_3=\sqrt{\frac{c_1^{\mathrm{NS5}}c_3^{\mathrm{NS5}}}{c_2^{\mathrm{NS5}}c_4^{\mathrm{NS5}}}}~.
\end{align}
The $\tau_i$'s are related to the complex-structure moduli, $U_i$,
and the dilaton, $S$, appearing in equation (\ref{2.18}) via
\cite{Camara:2005dc}:
\beal\label{complma}
S=\mathrm{e}^{-\phi_4}\frac{1}{\sqrt{{\tau_1}{\tau_2\tau_3}}};
~~~~~ U_1=\mathrm{e}^{-\phi_4}\sqrt{\frac{\tau_2\tau_3}{\tau_1}};
~~~~~U_2=\mathrm{e}^{-\phi_4}\sqrt{\frac{\tau_1\tau_3}{\tau_2}};
~~~~~U_3=\mathrm{e}^{-\phi_4}\sqrt{\frac{\tau_1\tau_2}{\tau_3}} ~.
\end{align}
In the above, $\phi_4$ is the four-dimensional dilaton. It is given by:
\beal\label{dilfour}
\mathrm{e}^{\phi_4}=\mathrm{e}^{\phi}\frac{1}{\sqrt{V}}~,
\end{align}
where $\phi$ is the ten-dimensional dilaton appearing in
(\ref{ktyb}), and $V$ is the volume of the $T^6$. The latter can
be read off in (\ref{ktyb}):
\beal\label{volu}
V=\left(c^{\mathrm{D8}} \right)^{-\frac{3}{2}}\left(\Pi_{\alpha=1}^4 c_{\alpha}^{\mathrm{NS5}} \right)^{\frac{3}{2}}
\sqrt{\Pi_{\beta=1}^3 c_{\beta}^{\mathrm{D4}} }~.
\end{align}
Putting (\ref{complma}, \ref{dilfour}, \ref{volu}) together we
arrive at:
\beal\label{complmb}
S&=\sqrt{c^{\mathrm{D8}}\Pi_{\alpha=1}^3 c_{\alpha}^{\mathrm{D4}}}
\sqrt{\frac{ c_{2}^{\mathrm{NS5}} c_{3}^{\mathrm{NS5}}c_{4}^{\mathrm{NS5}}}{c_{1}^{\mathrm{NS5}}} }\nn\\
U_1&=\sqrt{c^{\mathrm{D8}}\Pi_{\alpha=1}^3 c_{\alpha}^{\mathrm{D4}}}
\sqrt{\frac{ c_{1}^{\mathrm{NS5}} c_{3}^{\mathrm{NS5}}c_{4}^{\mathrm{NS5}}}{c_{2}^{\mathrm{NS5}}} }\nn\\
U_2&= \sqrt{c^{\mathrm{D8}}\Pi_{\alpha=1}^3 c_{\alpha}^{\mathrm{D4}}}
\sqrt{\frac{ c_{1}^{\mathrm{NS5}} c_{2}^{\mathrm{NS5}}c_{4}^{\mathrm{NS5}}}{c_{3}^{\mathrm{NS5}}} }\nn\\
U_3&=\sqrt{c^{\mathrm{D8}}\Pi_{\alpha=1}^3 c_{\alpha}^{\mathrm{D4}}}
\sqrt{\frac{ c_{1}^{\mathrm{NS5}} c_{2}^{\mathrm{NS5}}c_{3}^{\mathrm{NS5}}}{c_{4}^{\mathrm{NS5}}} }
~.
\end{align}

We are now ready to compare Eqs. (\ref{kahlm}, \ref{complmb}) with
the corresponding equation (\ref{solutionIIAg}), arrived at in the
effective-superpotential description. Upon taking (\ref{chrg})
into account, it can be seen that indeed there is perfect
agreement.

\subsubsection{D3/D5/D7/NS5/KK} \label{D3D5}

Let us now perform a T-duality along $x^1$. As already mentioned
in the introduction of Sec. \ref{ex}, this results in the
following flux superpotential:
\begin{equation}
 W_{\mathrm{IIB}}=i\tilde{e}_1U_1+
i\tilde{c}_2 U_2+i\tilde{c}_3 U_3+i\tilde{a}_0
S+i\tilde{c}_1T_1+i\tilde{e}_2 T_2+i \tilde{e}_3T_3+i\tilde{m}_0
U_1T_2T_3\, ,
\end{equation}
 where the last four terms are the geometrical-flux
contributions and guarantee all-moduli stabilization around the
type IIB $\mathrm{AdS}_4$ vacuum, as in the type IIA mirror
situation.

The corresponding brane configuration is now:
\bigskip
\begin{center}
  \begin{tabular}{|c||c|c|c|c|c|c|c|c|c|c|}
    \hline
    & $\xi^0$ & $\xi^1$  & $\xi^2$  & $y$ & $x^1$  & $x^2$  & $x^3$  & $x^4$  & $x^5$  & $x^6$ \\
    \hline
    \hline
    $\mathrm{D}3$ & $\bigotimes$ & $\bigotimes$  & $\bigotimes$ &  &  & $\bigotimes$  &   &
                &   &  \\
    \hline
    $\mathrm{D}5$ & $\bigotimes$ & $\bigotimes$  & $\bigotimes$ &   &  $\bigotimes$   &  & $\bigotimes$  &  $\bigotimes$
                 &   & \\
    \hline
    $\mathrm{D}5^{\prime}$ & $\bigotimes$ & $\bigotimes$  & $\bigotimes$ &  &$\bigotimes$   & &   &
                 & $\bigotimes$    &$\bigotimes$    \\

    \hline
    $\mathrm{D}7$ & $\bigotimes$ & $\bigotimes$  & $\bigotimes$ &   &   & $\bigotimes$ & $\bigotimes$    &$\bigotimes$
                 & $\bigotimes$  & $\bigotimes$ \\
\hline
    $\mathrm{NS5}$ & $\bigotimes$ & $\bigotimes$  & $\bigotimes$ &   &  $\bigotimes$  &  &  $\bigotimes$ &
                 & $\bigotimes$  & \\
\hline
    $\mathrm{NS5}^{\prime}$ & $\bigotimes$ & $\bigotimes$  & $\bigotimes$ &   &  $\bigotimes$  & &   &  $\bigotimes$
                 &  & $\bigotimes$ \\
 \hline
    $\mathrm{KK}$ & $\bigotimes$ & $\bigotimes$  & $\bigotimes$ &   &$\bullet$  & $\bigotimes$ &   &
               $\bigotimes$    &  $\bigotimes$  & \\
\hline
    $\mathrm{KK}^{\prime}$ & $\bigotimes$ & $\bigotimes$  & $\bigotimes$ &   & $\bullet$ & $\bigotimes$ &  $\bigotimes$     &
               &  &  $\bigotimes$ \\
\hline
   \end{tabular}
\end{center}
\bigskip
For simplicity, in deriving the explicit T-dualization of the
components of the bulk fields we will work in the near-horizon
region. After some rescalings of the coordinates of the six-torus,
the original type IIA solution can be written as:
\begin{align}
\mathrm{d}s_{10}^2&=
\mathrm{d}s^2_{\mathrm{AdS}_4}+\sum_{i=1}^6(\mathrm{d}x^i)^2 ;~~~\phi=\mathrm{const.};\nn\\
H_{x^2x^4x^6}&=H_{x^2x^3x^5}=H_{x^1x^3x^6}=H_{x^1x^4x^5}=a; \nn\\
F_{x^3x^4x^5x^6}&= F_{x^1x^2x^5x^6}=F_{x^1x^2x^3x^4}=b; ~~~F=c ~,\nn
\end{align}
where $a$, $b$,  $c$ are constants, and $b=3/5c$.
Upon T-dualising along $x^1$ (we
follow the T-duality rules as given in \cite{hassan}) we obtain the
following IIB solution:
\begin{align}\label{iibsol}
\mathrm{d}s_{10}^2&=
\mathrm{d}s^2_{\mathrm{AdS}_4}+\sum_{i=2}^6(\mathrm{d}x^i)^2 +(\mathrm{d}x^1-A)^2;~~~\phi=\mathrm{const.};\nn\\
 F_{1}&=c(\mathrm{d}x^1-A);~~~
F_{3}=b (\mathrm{d}x^2\wedge \mathrm{d}x^3\wedge \mathrm{d}x^4
+\mathrm{d}x^2\wedge \mathrm{d}x^5\wedge \mathrm{d}x^6); \nn\\
H_{3}&= a(\mathrm{d}x^2\wedge \mathrm{d}x^4\wedge \mathrm{d}x^6
+\mathrm{d}x^2\wedge \mathrm{d}x^3\wedge \mathrm{d}x^5);
~~~F_{x^1x^3x^4x^5x^6}=b~,
\end{align}
where all other flux components vanish, except for the ``external"
component of the five-form Ramond flux $F_{\xi^0\xi^1\xi^2 y x^2}$
which is determined from the one above by self-duality. These
fluxes correspond precisely to the brane configuration presented
in the table above.

The one-form connection $A$ is given by:
\beal
A=a(x^6\mathrm{d}x^3+x^5\mathrm{d}x^4)~.
\end{align}
Hence $x^1$ should be thought of as the coordinate of a twisted
$S^1$ fibre over a $T^5$ base parameterized by $x^2,\ldots,x^6$.
It follows that the geometry of the internal six-dimensional space
is that of a {\em twisted torus}, or a {\em nilmanifold}. More
precisely: defining left-invariant one-forms
$\mathrm{e}^1:=\mathrm{d}x^1-A$ and $\mathrm{e}^i:=\mathrm{d}x^i$
for $i={2,\ldots ,6}$, it is straightforward to see that the
internal space can be identified with the nilmanifold 5.1 of table 4
of \cite{scan}. To make contact with the T-duality rules for NS5
branes, one may think of the internal nilmanifold as consisting of
two intersecting KK monopoles (which are purely gravitational
backgrounds) \cite{grossperry} along the directions indicated with
bullets in the table above. Indeed, after T-duality, the
NS5-branes are traded for Kaluza--Klein monopoles (TAUB-NUT
spaces), carrying NUT-charges in the marked directions.

From (\ref{iibsol}) and the source-modified Bianchi identities
$\mathrm{d}F+H\wedge F=j$ we can read off the following non-zero
source components:
\beal
j^{\mathrm{O5}}&=ac(\mathrm{d}x^2\wedge \mathrm{d}x^4\wedge \mathrm{d}x^6
+\mathrm{d}x^2\wedge \mathrm{d}x^3\wedge \mathrm{d}x^5)\wedge (\mathrm{d}x^1-A)\nn\\
j^{\mathrm{O7}}&=ac(\mathrm{d}x^3\wedge \mathrm{d}x^6
+\mathrm{d}x^4\wedge \mathrm{d}x^5)~.
\end{align}
Hence we have in addition O5/O7 orientifold planes along:
\bigskip
\begin{center}
  \begin{tabular}{|c||c|c|c|c|c|c|c|c|c|c|}
    \hline
    & $\xi^0$ & $\xi^1$  & $\xi^2$  & $y$ & $x^1$  & $x^2$  & $x^3$  & $x^4$  & $x^5$  & $x^6$ \\
    \hline
    \hline
    $\mathrm{O}5$ & $\bigotimes$ & $\bigotimes$  & $\bigotimes$ & $\bigotimes$ &  &  & $\bigotimes$   &
               &    $\bigotimes$ &  \\

    \hline
    $\mathrm{O}5^{\prime}$ & $\bigotimes$ & $\bigotimes$  & $\bigotimes$ &   $\bigotimes$ &  &  & &
               $\bigotimes$  &   &    $\bigotimes$\\

    \hline
    $\mathrm{O}7$ & $\bigotimes$ & $\bigotimes$  & $\bigotimes$ &    $\bigotimes$ &     $\bigotimes$ &  $\bigotimes$ &   &
                 $\bigotimes$  & $\bigotimes$  &  \\
\hline
    $\mathrm{O}7^{\prime}$ & $\bigotimes$ & $\bigotimes$  & $\bigotimes$ &   $\bigotimes$  &  $\bigotimes$  &
 $\bigotimes$ &  $\bigotimes$ &
                 &   &   $\bigotimes$\\
\hline
   \end{tabular}
\end{center}
\bigskip
This is precisely what one would expect by applying one T-duality
along $x^1$ to the O6-planes of the model generated by the
D4/D8/NS5-branes discussed in Sec. \ref{tabela}.

Performing another T-duality along $x^2$ brings us back to type
IIA. Now all NS5-branes have been replaced by KK monopoles and the
sources are D2/D6/KK-branes/monopoles distributed as
\bigskip
\begin{center}
  \begin{tabular}{|c||c|c|c|c|c|c|c|c|c|c|}
    \hline
    & $\xi^0$ & $\xi^1$  & $\xi^2$  & $y$ & $x^1$  & $x^2$  & $x^3$  & $x^4$  & $x^5$  & $x^6$ \\
    \hline
    \hline
    $\mathrm{D}2$ & $\bigotimes$ & $\bigotimes$  & $\bigotimes$ &  &  &   &   &
                &   &  \\
    \hline
    $\mathrm{D}6$ & $\bigotimes$ & $\bigotimes$  & $\bigotimes$ &   &  $\bigotimes$   & $\bigotimes$   & $\bigotimes$  &  $\bigotimes$
                 &   & \\
    \hline
    $\mathrm{D}6^{\prime}$ & $\bigotimes$ & $\bigotimes$  & $\bigotimes$ &  &$\bigotimes$   &$\bigotimes$  &   &
                 & $\bigotimes$    &$\bigotimes$    \\

    \hline
    $\mathrm{D}6^{\prime\prime}$ & $\bigotimes$ & $\bigotimes$  & $\bigotimes$ &   &   &  & $\bigotimes$    &$\bigotimes$
                 & $\bigotimes$  & $\bigotimes$ \\
\hline
    $\mathrm{KK}$ & $\bigotimes$ & $\bigotimes$  & $\bigotimes$ &   &  $\bigotimes$  &  $\bullet$ &  $\bigotimes$ &
                 & $\bigotimes$  & \\
\hline
    $\mathrm{KK}^{\prime}$ & $\bigotimes$ & $\bigotimes$  & $\bigotimes$ &   &  $\bigotimes$  & $\bullet$ &   &  $\bigotimes$
                 &  & $\bigotimes$ \\
 \hline
    $\mathrm{KK}^{\prime\prime}$ & $\bigotimes$ & $\bigotimes$  & $\bigotimes$ &   &$\bullet$  & $\bigotimes$ &   &
               $\bigotimes$    &  $\bigotimes$  & \\
\hline
    $\mathrm{KK}^{\prime\prime\prime}$ & $\bigotimes$ & $\bigotimes$  & $\bigotimes$ &   & $\bullet$ & $\bigotimes$ &  $\bigotimes$     &
               &  &  $\bigotimes$ \\
\hline
   \end{tabular}
\end{center}
\bigskip
The superpotential receives contributions from $F_2$ and $F_6$
Ramond fields as well as from geometrical fluxes. We will not
elaborate on this model here.

\subsubsection{D4/D8/NS5 and D5/NS5}

We would like finally to consider the situation where instead of
four stacks of  NS5-branes, we keep only one in the model
presented in Sec. \ref{ddno}:
\bigskip
\begin{center}
  \begin{tabular}{|c||c|c|c|c|c|c|c|c|c|c|}
    \hline
    & $\xi^0$ & $\xi^1$  & $\xi^2$  & $y$ & $x^1$  & $x^2$  & $x^3$  & $x^4$  & $x^5$  & $x^6$ \\
    \hline
    \hline
    $\mathrm{D}4$ & $\bigotimes$ & $\bigotimes$  & $\bigotimes$ &  &$\bigotimes$  &  $\bigotimes$ &   &
                &   &  \\

    \hline
    $\mathrm{D}4^{\prime}$ & $\bigotimes$ & $\bigotimes$  & $\bigotimes$ &  &  &  & $\bigotimes$  &
               $\bigotimes$  &   &  \\

    \hline
    $\mathrm{D}4^{\prime\prime}$ & $\bigotimes$ & $\bigotimes$  & $\bigotimes$ &   &   &  &   &
                 & $\bigotimes$  & $\bigotimes$ \\
\hline
    $\mathrm{NS}5$ & $\bigotimes$ & $\bigotimes$  & $\bigotimes$ &   &  $\bigotimes$  &   &  $\bigotimes$ &
                 & $\bigotimes$  & \\
 \hline
    $\mathrm{D}8$ & $\bigotimes$ & $\bigotimes$  & $\bigotimes$ &   &  $\bigotimes$ & $\bigotimes$ & $\bigotimes$  &  $\bigotimes$
                 & $\bigotimes$  & $\bigotimes$ \\
\hline
   \end{tabular}
\end{center}
\bigskip
This intersecting-brane distribution corresponds to the IIA
superpotential Eq. (\ref{IIA}) with no complex-structure moduli.
Its supergravity brane solution is simply obtained from
(\ref{ktyb}), by identifying all four NS5 harmonic functions, i.e.
setting all $c_\alpha^{\rm NS}$ resp. $d_\alpha^{\rm NS}$ to equal
values.

The type IIB mirror brane set up of three stacks of D4-branes, one
of D8-branes and one of NS5-branes is obtained by performing a
T-duality in three directions $(x^1,x^3,x^5)$:
\bigskip
\begin{center}
  \begin{tabular}{|c||c|c|c|c|c|c|c|c|c|c|}
    \hline
    & $\xi^0$ & $\xi^1$  & $\xi^2$  & $y$ & $x^1$  & $x^2$  & $x^3$  & $x^4$  & $x^5$  & $x^6$ \\
    \hline
    \hline
    $\mathrm{D}5$ & $\bigotimes$ & $\bigotimes$  & $\bigotimes$ &  &  & $\bigotimes$ &   $\bigotimes$&
                &  $\bigotimes$ &  \\

    \hline
    $\mathrm{D}5^{\prime}$ & $\bigotimes$ & $\bigotimes$  & $\bigotimes$ &  & $\bigotimes$ &  &  &
               $\bigotimes$  &$\bigotimes$   &  \\

    \hline
    $\mathrm{D}5^{\prime\prime}$ & $\bigotimes$ & $\bigotimes$  & $\bigotimes$ &   &  $\bigotimes$ &  &   $\bigotimes$&
                 &   & $\bigotimes$ \\
                 \hline
    $\mathrm{D}5^{\prime\prime\prime}$ & $\bigotimes$ & $\bigotimes$  & $\bigotimes$ &   &   & $\bigotimes$ & &  $\bigotimes$
                 &   & $\bigotimes$ \\
\hline
    $\mathrm{NS}5$ & $\bigotimes$ & $\bigotimes$  & $\bigotimes$ &   &  $\bigotimes$ &   &  $\bigotimes$ &
                 & $\bigotimes$  & \\
 \hline
     \end{tabular}
\end{center}
\bigskip
This brane set up corresponds to the IIB superpotential
(\ref{tv1}), which does not depend on K\"ahler moduli. Here,
however, the model is not the non-geometrical one discussed in
Secs. \ref{2.2} and \ref{2.3}, and the Kähler moduli are not
stabilized.

\section{Conclusions}\label{sec:conc}

Flux compactifications have opened ways to compute macroscopically
the entropy of gravitational backgrounds. The analysis of the
source content of these backgrounds provides a microscopic
perspective for this important issue.

With these motivations in mind, we have here explored the
appearance of $\mathrm{AdS}_4$ vacua in flux compactifications.
The macroscopic approach is based on the effective superpotential
of gauged supergravity, which is a polynomial in the chiral
superfields with coefficients directly related to the flux
numbers. In the absence of geometrical fluxes induced by the
Scherk--Schwarz mechanism, complete moduli stabilization can be
achieved in type IIA models, whereas in perturbative type IIB this
is possible only in non-geometric compactifications where Kähler
moduli are absent. It should however be stressed that geometrical
fluxes are not a problem \emph{per se}; they simply enforce a
generalized, torsionful structure.

An important result is the analysis of the microscopic origin
of the above $\mathrm{AdS}_4$ vacua. The latter appear as
near-horizon geometries of brane/monopole distributions. The most
interesting type IIA construction is based on a D4/D8/NS5-brane
configuration, where some of the branes are smeared. This latter property plays a crucial role,
and allows in particular to evade the introduction of geometrical
fluxes in the macroscopic, gauged-supergravity picture. From the
microscopic point-of-view all consistency conditions remain
satisfied, provided the various sources and charges are carefully
taken into account, and orientifold planes added appropriately.

As we saw in Sec. \ref{ddno}, zooming in on the near-horizon
region of the D4/D8/NS5 intersecting-brane solution amounts to
replacing the D4/D8/NS5 branes with fluxes, while adding a set of
completely smeared orientifold six-planes. This corresponds to a
consistent ten-dimensional supergravity picture: it is contained
in the class of solutions of \cite{Lust:2004ig} modified by the
addition of smeared O6-planes. Such models were considered
recently in detail by Acharya \emph{et al.} \cite{acha}.

A word of caution is in order: as we have seen in Sec. \ref{ddno},
in the near-horizon limit the charge distributions corresponding
to our brane setup are uniquely determined by the requirement that
the near-horizon geometry be $\mathrm{AdS}_4\times T^6$ and the
dilaton approach a finite value. However, the subleading behaviour
is largely arbitrary. It is fair to say that this behaviour should
be better understood, and a physical interpretation should be
provided, before a fully satisfactory picture can be said to have
been given.

Upon T-dualizing type IIA models, new type IIB setups emerge. An
interesting situation is reached in this way when a T-duality is
performed on the above  D4/D8/NS5-brane plus O6-plane system. Some of the
NS5-branes become Kaluza--Klein monopoles (sources for geometric
fluxes), which are purely gravitational backgrounds. The type IIB
$\mathrm{AdS}_4$ vacuum appears then as the near-horizon geometry
of a D3/D5/D7/NS5/KK brane/monopole distribution. Going to the near-horizon
region of the D3/D5/D7/NS5/KK system amounts to replacing the
branes by their fluxes while adding a system of smeared O5/O7
planes.

The price for the moduli stabilization around this IIB
$\mathrm{AdS}_4$ vacuum is that the internal six-dimensional
geometry is now a certain nilmanifold. This model is therefore a
concrete realization of general ideas about flux compactifications
and generalized geometries, and is an important example
independently of the issues of moduli stabilization. Analyzing the
emergence of spaces with generalized $G\times G$ group structure
by using the tools presented in this paper, i.e. in direct contact
with four-dimensional physics, deserves further attention.

Let us mention that, as already noted, there is a certain
ambiguity in the interpretation of the smeared sources in the
context of  supergravity. One way to distinguish between the
different possible interpretations would be to examine the
next-to-leading order derivative corrections to our solutions;
indeed, D-branes and orientifolds behave differently beyond
leading order in the $\alpha'$ expansion \cite{orie}.

Coming back to our original motivation, we would like to add that
it is not straightforward to assign macroscopic entropy to an
$\mathrm{AdS}_4$ space, mostly because we do not know of any
thermodynamical system that would support such an entropy. We
could nevertheless use the dual, brane-source picture. The
set of microscopic degrees of freedom certainly allows one to define
an entropy, but this does not necessarily shed any light
on the other (macroscopic) side of the duality: it merely serves as a definition.
We hope to investigate these interesting directions in the future.

\vskip0.5cm

\begin{acknowledgments}
  The authors would like to thank  Gabriel Lopes Cardoso, Paul Koerber and Luca Martucci
  for stimulating discussions. Marios Petropoulos thanks the Arnold-Sommerfeld-Center for
  Theoretical Physics of the LMU-München and the MPI-München for kind hospitality and financial support.
  This work was supported in part by the EU under the contracts MEXT-CT-2003-509661,
  MRTN-CT-2004-005104, MRTN-CT-2004-503369 and by the Agence Nationale pour la
  Recherche, France, contract 05-BLAN-0079-01.
\end{acknowledgments}

\appendix

\section{Supersymmetry}\label{appsus}

The purpose of this appendix is to provide further details of the
supersymmetry of solution (\ref{ktyb}) of Sec. \ref{ddno}.

The calibration
conditions for the branes follow from the kappa symmetry of the
world-volume action, which, in the doubly-supersymmetric approach, can be thought of as
a gauge-fixing of supersymmetry. In the case of the static, flat branes without
world-volume excitations, considered here, the calibration conditions \cite{gencalold}
take the following standard form (see e.g. \cite{gaun, youm}):
\beal
\Gamma_{\xi^0\xi^1\xi^2 x^{i_1}\dots x^{i_{p-2}} }\epsilon_+=\epsilon_-~,
\label{calal}
\end{align}
for RR $p$-branes, and
\beal
\Gamma_{\xi^0\xi^1\xi^2 x^{i}x^j x^{k} }\epsilon=\epsilon~,
\label{calbe}
\end{align}
for type IIA NS5-branes. In (\ref{calal}) the $p$-brane is taken
along the $\xi^0, \xi^1, \xi^2, x^{i_1}, \dots, x^{i_{p-2}}$
directions; in (\ref{calbe}) the NS5-brane is along $\xi^0, \xi^1,
\xi^2, x^{i}, x^j, x^{k}$. The products of gamma-matrices above
are in an orthonormal frame; $\Gamma_{11}$ is the chirality
operator in ten dimensions, so that:
\beal
\epsilon_{\pm}=\frac{1}{2}(1\pm \Gamma_{11})~\epsilon~,
\end{align}
where $\epsilon$ is the ten-dimensional Killing spinor. The latter can be taken to be constant,
up to a $y$-dependent conformal factor, where $y$ is the overall transverse coordinate.

Conditions (\ref{calal}) and (\ref{calbe}) have to be imposed
individually for each brane in the system. A priori, each
condition reduces the supersymmetry by one-half, but not all
conditions are necessarily independent. In the example of Sec.
\ref{ddno}, there are four D-branes and four NS5-branes, and
therefore eight conditions following from (\ref{calal}),
(\ref{calbe}). However, it can easily be seen that only four of
them are independent; they can be taken as in Eq. (\ref{calmo}).

The most straightforward way to check supersymmetry is then to
simply substitute the values of all fields in (\ref{ktyb}) into
Eqs. (\ref{susytr}), taking (\ref{calmo}) into account.
Alternatively, if the reader prefers to use the formalism of
Romans' supergravity \cite{roma}, the vielbein has to be rescaled
from $e_m{}^a$ in the string frame to
$\mathrm{e}^{-\frac{\phi}{2}}e_m{}^a$ in the Einstein frame, where
$\phi$ is the dilaton. Moreover, the following dictionary between
the democratic and the Romans' formalism (in the conventions of
e.g. \cite{Lust:2004ig}) has to be used:
\medskip
\begin{center}
  \begin{tabular}{|c|c|}
    \hline
    Democratic & Romans \\
    \hline
    \hline
   $F_{0}$ &  $-2m$ \\

    \hline
     $F_{2}$ &  $-2mB^{\prime}$\\

    \hline
   $F_{4}$ & $-G$ \\
                 \hline
    $H_{3}$ & $-H$ \\
 \hline
     \end{tabular}
\end{center}
\medskip

\end{document}